\documentclass{aa}
\usepackage{graphicx}
\usepackage{subcaption}
\usepackage{xcolor}
\usepackage[colorlinks]{hyperref}
\usepackage{subdepth}  
\usepackage{txfonts}
\usepackage{stmaryrd}
\usepackage{float}

\newcommand*{\vv}[1]{\boldsymbol{#1}}
\newcommand{\ud}{\,\mathrm{d}}
\newcommand{\ft}[1]{\,\mathcal{F}\!\left[ #1 \right]}
\newcommand{\invft}[1]{\,\mathcal{F}^{\scriptscriptstyle -1}\!\left[ #1 \right]}

\begin{document}

   \title{Single-mode interferometric field of view\\in partial turbulence correction}

   \subtitle{Application to the observation of the environment of Sgr A* with GRAVITY}

   \author{G. Perrin\inst{1}
          \and
          J. Woillez\inst{2}
          }

   \institute{LESIA, Observatoire de Paris, Universit\'e PSL, CNRS, Sorbonne Universit\'e, Univ. Paris Diderot, Sorbonne Paris Cit\'e, 5 place Jules Janssen, 92195 Meudon, France.\\
             \email{guy.perrin@obspm.fr}
         \and
             European Southern Observatory, Karl-Schwarzschild-Str. 2, 85748 Garching bei München, Germany.\\
             \email{jwoillez@eso.org}
             }

   \date{Received 3 August 2018; accepted 18 March 2019}

  \abstract
   {The GRAVITY instrument on the ESO VLTI is setting a new mark in the landscape of optical interferometers. Long exposures are possible for the first time in this wavelength domain, delivering a dramatic improvement for astrophysics. In particular, faint objects can be observed at the angular resolution of the VLTI, with exposures exceeding by many orders of magnitude the coherence time of atmospheric turbulence.}
   {Modern interferometers, especially those that combine light collected by large telescopes such as the Unit Telescopes of the VLT, benefit from partial correction of atmospheric turbulence. We investigate in this paper the influence of atmospheric turbulence on the maximum field of view of interferometers such as GRAVITY, where wavefronts are filtered with single-mode fibres. The basic question is whether the maximum field of view is restricted to the diffraction limit of single apertures or if it is wider in practice. We discuss in particular the case of the field around Sgr A* , with an emphasis on the detectability of the massive main-sequence star S2.}
   {We theoretically investigated the interferometric and photometric lobes of the interferometer and took atmospheric turbulence into account. We computed the lobe functions with and without partial correction for atmospheric turbulence.}
   {The main result of the paper is that the field of view of the interferometer is widened by tip-tilt residues if higher orders of atmospheric turbulence are corrected for. As a particular case, the S2 star can be detected in interferometric frames centred on Sgr A* even when the distance between the two objects is up to about twice the diffraction limit of a single aperture. We also show that the visibilities are not biased in this case if the pointing accuracies of the fibres are of the order of 10\,mas.}
   {}

   \keywords{instrumentation: high angular resolution --instrumentation: interferometers -- instrumentation: adaptive optics -- atmospheric effects -- Galaxy: center}

   \maketitle
%

\section{Introduction}

The question of the field of view in optical-infrared interferometry is still a matter of investigation.
It was long thought that the question of the field of view of an interferometer was connected to the pupil configuration.
\citet{Traub1986} stated the golden rule of interferometry: the maximum field of view would be preserved in an interferometer if the entrance and exit pupils were homothetic because in this case the interference pattern is the result of the convolution of the object by the interferometer response.
\citet{Labeyrie1996} proposed the hypertelesope concept, where the exit pupil is densified relative to the entrance pupil. The convolution relation is lost in this case, as is also the case for most interferometers.
The exit pupil is always densified in practice for reasons of the signal-to-noise ratio when limited by detector noise because this minimises the number of fringes across the diffraction pattern.
\citet{Lardiere2007} showed that even in the case of pupil densification, the field of view does not so much depend on the exit pupil, but is given by the smallest separation in the array $s$: $\lambda/s$. Direct imaging can be achieved by a hypertelescope in a relatively large field of view.

Beyond array and beam combiner geometries, turbulence also plays a role in interferometric imaging field of view.
\citet{RoddierLena1984a,RoddierLena1984b} investigated the effect of turbulence in both the single-speckle and multi-speckle interferometry cases for objects that are not resolved by a single aperture.
In particular, they derived the expression of the coherence term measured by the Fizeau-type (multi-axial) interferometer.
\citet{Mourard+1994} later refined this estimator to improve the stability and fidelity of the measured visibilities.
Despite these achievements, visibility accuracies reached a few percent to a percent at best.
Following earlier ideas to use single-mode fibres to overcome the difficulty of calibrating visibilities in atmospheric turbulence conditions \citep{Froehly1981}, the FLUOR concept was proposed \citep{CoudeDuForesto1992}.
\citet{CoudeDuForesto+2000} investigated the coupling of starlight to a single-mode fibre at the focus of a telescope under partial correction for wavefront turbulence with adaptive optics.
For maximum coupling to the fibre without aberrations, the mode of the fibre has to match the diffraction limit $\lambda/D$ of each aperture.
In presence of turbulence, the injection efficiency is proportional to the Strehl ratio or coherent energy, as demonstrated with simulations by \citet[][   ]{Haubois2009}.
\citet{Guyon2002} investigated the effect of turbulence in the short-exposure case on the field of view of a single-mode interferometer: the interferometer measures the object that is apodised by the fibre mode.
He showed that calibration over the whole field of view would be difficult, however.
In parallel, \citet{Mege2002} built a theoretical framework for the measurement of coherence with single-mode interferometers.
He investigated the field of view of a single-mode interferometer and showed that it is limited by the fibre lobe in absence of turbulence and static aberrations or with turbulence in the long-exposure case but without correction of turbulence.
Most of these findings (without turbulence) were known from radio astronomy, where light is also filtered by single-mode waveguides \citep{Thompson+2017}.

We take the next step in these studies and investigate in this paper the influence of partial correction for turbulent wavefronts on the field of view of a single-mode interferometer.
We first describe the field of view that is transported by a single-mode waveguide in Section~\ref{Sec:2}.
In Section~\ref{Sec:3} we derive the expression of the instantaneous visibility measured by a single-mode interferometer.
This expression is generalised to the long-exposure case without any phase correction in Section~\ref{Sec:4}.
It is further discussed in Section~\ref{Sec:5}, when a partial correction for atmospheric turbulence is applied, with the consequence on the field of view of the interferometer.
The overall results are discussed in Section~\ref{Sec:6}
 

\begin{figure}[t]
    \includegraphics[width=\linewidth]{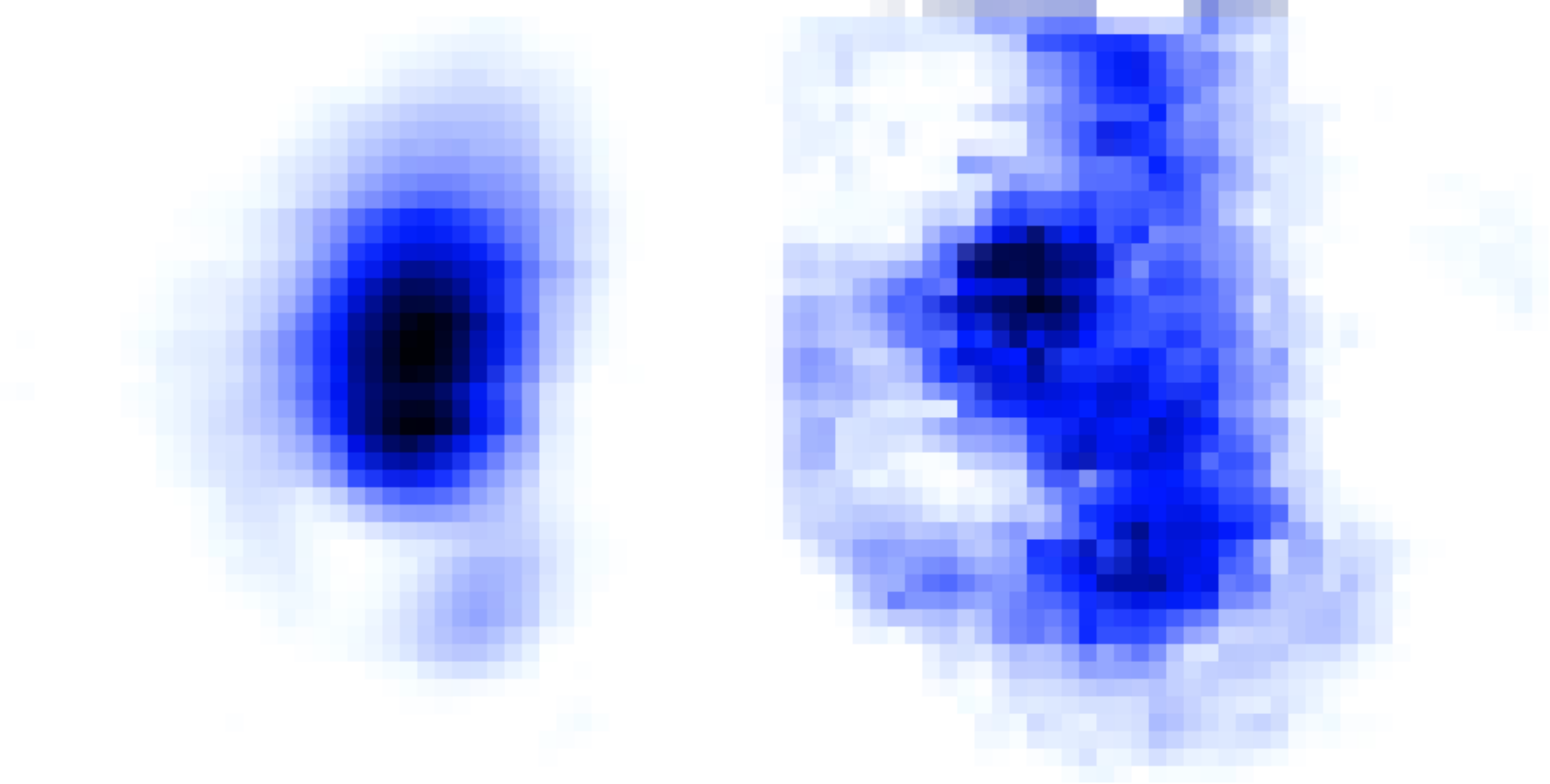}
    \caption{Telescope lobe measured on an internal source (left) and on a star with partial adaptive optics correction (right), obtained with the `OHANA experiment \citep{Perrin+2006} on one of the telescopes of the Keck Interferometer \citep{Colavita+2013}.}
    \label{Fig:lobe}
\end{figure}

\section{Field of view transported by a single-mode waveguide}
\label{Sec:2}

The coupling of the light collected by a single telescope into a single-mode fibre is set by the overlap integral $\eta$ between the field at the telescope focus $E_\mathrm{tel}$ and the fibre mode $E_\mathrm{fib}$ \citep{Neumann1988}.
The wave $E(\vv\beta)$ injected into the fibre then writes 
  \begin{equation}
    E(\vv{\beta}) = E_\mathrm{fib}(\vv\beta) \times \int E_\mathrm{tel}(\vv\xi) \,  E^{\star}_\mathrm{fib}(\vv\xi) \ud\vv\xi = E_\mathrm{fib}(\vv\beta) \times \eta,
    \label{Eq:OverlapIntegral}
  \end{equation}
where $\vv\beta$ is the local coordinates against which the fibre mode $E_\mathrm{fib}$ is expressed, and assuming a normalised fibre mode profile, $\int |E_\mathrm{fib}(\vv\beta)|^2 \ud\vv\beta = 1$.
When the field at the telescope focus changes, only the global amplitude of the injected wave changes, not the profile $E_\mathrm{fib}(\vv\beta)$ of the fibre field, also called carrier wave.
This is the mathematical expression of the spatial filtering properties of a single-mode fibre.
Through the Parseval-Plancherel theorem, the overlap integral can be converted into pupil plane as follows:
  \begin{equation}
    \eta = \int{ \invft{E_\mathrm{tel}}(\vv{u}) \, \invft{E_\mathrm{fib}}^\star(\vv{u}) \, \ud\vv{u} },
  \end{equation}
where $\invft{}$ denotes the inverse Fourier transform.
The inverse Fourier transform of the focused wave, $\invft{ E_\mathrm{tel} }$, is the complex telescope pupil $P$ multiplied by the wave $W$ collected by the telescope (with $W=\invft{ E_\mathrm{obj} }$, where $E_\mathrm{obj}$ is the wave from the object in image plane).
The inverse Fourier transform of the quasi-Gaussian fibre mode, $\invft{ E_\mathrm{fib} }$, acts as a quasi-Gaussian apodising function in pupil plane.
Defining $P_\varocircle$, the complex pupil apodised by the fibre mode,
  \begin{equation}
    P_\varocircle(\vv{u}) = P(\vv{u}) \, \invft{ E_\mathrm{fib} }^{\star} (\vv{u}),
    \label{Eq:ApodisedPupil}
  \end{equation}
the overlap integral becomes
  \begin{equation}
    \eta = \int P_\varocircle(\vv{u}) W(\vv{u}) \ud\vv{u},
  \end{equation}
an average of the collected wave weighted by the complex apodised pupil.

The intensity associated with the coupled field relates to the overlap integral as follows:
  \begin{equation}
    I = \int \left<\left| E(\vv\beta) \right|^2\right> \ud \vv{\beta} = \left<\left| \eta \right|^2\right>,
    \label{Eq:TelescopeLobe}
  \end{equation}
where $\left<\right>$ denotes the ensemble average operator on the oscillations of the wave $W$.
At optical and infrared wavelengths, the variation timescale of the apodised complex pupil functions $P_\varocircle$ as a result of atmospheric turbulence or vibrations, for instance, is much longer than the period of the waves.
The ensemble average can be approximated with good accuracy by an average over many cycles of the wave, while keeping the averaging time shorter than the coherence time of the apodised complex pupil functions fluctuations.
Doing so, the intensity becomes
  \begin{equation}
    I = \iint P_\varocircle(\vv{u}) P_\varocircle^\star(\vv{v}) \left< W(\vv{u}) W^\star(\vv{v}) \right> \ud\vv{u} \ud\vv{v}.
  \end{equation}
Recognising the complex visibility of the object $V(\vv{u}-\vv{v})$ in the term $\left<W(\vv{u}) W^\star(\vv{v})\right>$, assuming the integral of the object intensity is normalised: $\int O(\vv\alpha) \ud\vv\alpha=1$, where $O(\vv\alpha)=| E_\mathrm{obj} |^2 (\vv\alpha)$, and applying a $\vv{w}=\vv{v}-\vv{u}$ coordinates change, the previous equation can be re-written as follows:
  \begin{equation}
    I = \int {  \left[ P_\varocircle \otimes P_\varocircle \right] (\vv{w}) V(\vv{w}) \ud\vv{w}},
    \label{Eq:SingleVisibilityAveraging}
  \end{equation}
where $P_\varocircle \otimes P_\varocircle$ is the auto-correlation of the apodised pupil.
With the Perseval-Plancherel theorem used once more, the equation above can be converted into object space,
  \begin{equation}
    I = \int \invft{P_\varocircle \otimes P_\varocircle}(\vv\alpha) O(\vv\alpha) \ud\vv\alpha
  ,\end{equation}
where the inverse Fourier transform of the visibility was identified as the observed object $O$.

From the equation above, the inverse Fourier transform of the apodised pupil auto-correlation
  \begin{equation}
    L(\vv\alpha) = \invft{P_\varocircle \otimes P_\varocircle}(\vv\alpha)
    \label{Eq:PhotometricLobe}
  \end{equation}
is recognised as the single-mode telescope lobe $L$: the map of the intensity coupled to the fibre as a function of off-axis distance.
This is equivalent to the point spread function of incoherent Fraunhofer imaging.
Finally, to prepare a comparison with the single-mode coherence measurement of the next section, the intensity coupled into the fibre can also be presented as
  \begin{equation}
    I = \ft{LO}(\vv{0}),
    \label{Eq:Photometry-Lobe}
  \end{equation}
the zero-frequency visibility of the object apodised by the single-mode telescope lobe.

This result is similar to the well-known effect of the antenna lobe of radio telescopes \citep{Thompson+2017} and was also established by \citet{Mege2002}.
The apodised pupil $P_\varocircle$, just like its non-apodised counter part, is a complex quantity.
Its non-zero phase is the result of static aberrations and dynamic turbulence, or the residual phase if an adaptive optics system is present on the telescope.
When the strength of the residual turbulence increases, the autocorrelation of the complex apodised pupil narrows and the telescope lobe widens.
Then, the coupled intensity becomes more contaminated by surrounding objects when the turbulence is strong.
An image of the coupling map can be produced by moving the fibre in the image plane or by stirring a mirror in pupil plane to scan the $\vv{\alpha}$ directions.
This is how the telescope lobes illustrated in Fig.~\ref{Fig:lobe} were obtained with the 'OHANA experiment on the Keck telescopes \citep{Perrin+2006}; it is affected by static aberrations and residual atmospheric turbulence.

\section{Instantaneous visibility measured by single-mode interferometers}
\label{Sec:3}

We consider $E_i(\vv\beta)$, the wave injected in the fibre at the focus of telescope $i$.
The fields $E_1(\vv\beta)$ and $E_2(\vv\beta)$ from two telescopes overlap in the waveguide after beam combination \citep[e.g. with an X-coupler,][]{Jocou+2012}, with the resulting field
  \begin{equation}
    E_{12}(\vv{\beta}) = E_1(\vv\beta) + E_2(\vv\beta),
  \end{equation}
where for the sake of simplicity, the complex coefficients of the beam combiner were chosen equal to unity.
The output of the waveguide is conjugated with the detector in a focal plane where the intensity $I_{12}$ is averaged over $\vv{\beta}$:
  \begin{equation}
    I_{12} = \int{ \left< \left| E_{12}(\vv{\beta})  \right|^2 \right> \ud\vv\beta },
    \label{Eq:Interferogram}
  \end{equation}
where $\left< \right>$ denotes the same ensemble average operator as in section~\ref{Sec:2}.
Developing the square leads to the definition of the complex coherence factor $\gamma_{12}$ measured between single-mode telescopes 1 and 2, which relates to the overlap integrals $\eta_1$ and $\eta_2$ for each telescope as follows:
  \begin{equation}
    \gamma_{12}=\int{\left< E_1(\vv{\beta})\, E_2^\star(\vv{\beta}) \right>  \ud\vv{\beta}} = \left< \eta_1 \eta_2^\star \right>.
    \label{Eq:Coherence1}
  \end{equation}
The equation above is the two-telescope equivalent of Eq.~\ref{Eq:TelescopeLobe}.
The same transformations as in the previous section yields a two-telescope equivalent of Eq.~\ref{Eq:SingleVisibilityAveraging}, linking the complex coherence factor to the object visibility:
  \begin{equation}
    \gamma_{12} = \int \left[ P_{\varocircle,1} \otimes P_{\varocircle,2} \right] (\vv{u}) V(\vv{u}+\vv{B}/\lambda) \ud\vv{u},
    \label{Eq:DualVisibilityAveraging}
  \end{equation}
where both apodised pupils $P_{\varocircle,1}$ and $P_{\varocircle,2}$ are centred on the origin, and $\vv{B}$ represent the baseline vector joining the two pupil centres.
Through the Parseval-Plancherel theorem, the complex coherence factor expressed in object space writes
\begin{equation}
    \gamma_{12} = 
    \int \invft{P_{\varocircle,1} \otimes P_{\varocircle,2}} (\vv\alpha) O(\vv\alpha) e^{-i2\pi \vv{B}/\lambda\cdot\vv\alpha}  \ud\vv\alpha
    \label{Eq:19}
.\end{equation}
By defining the single-mode interferometric lobe $L_{12}$ as the inverse Fourier transform of the apodised pupil cross-correlation,
  \begin{equation}
    L_{12}(\vv\alpha) = \invft{P_{\varocircle,1} \otimes P_{\varocircle,2}} (\vv\alpha),
    \label{Eq:InterferometricLobe}
  \end{equation}
the complex coherence factor writes  \begin{equation}
    \gamma_{12} = \ft{L_{12}O}(\vv{B}/\lambda).
    \label{Eq:Cohrence-Lobe}
  \end{equation}

The coherence measured by the single-mode interferometer is also the visibility function of the object that is apodised by the instantaneous interferometric lobe.
\citet{Mege2002} obtained a similar result.
In the absence of turbulence, this expression is valid regardless of exposure time, as long as it is much longer than the period of the wave.
In this particular case, in the absence of any static or turbulent aberration and with identical telescopes, all three apodised pupils $P_{\varocircle,1}$, $P_{\varocircle,2}$, and $P_\varocircle$ are identical, and so are the lobes $L_{12}(\vv\alpha) = L(\vv\alpha)$.
The maximum field of view of the interferometer is then set by the pupils that are apodised by the mode of the fibres.

\section{Single-mode fully turbulent long exposure}
\label{Sec:4}

In this section, wavefronts at the telescopes are affected by atmospheric turbulence with Gaussian statistics, and no adaptive optics or fringe tracking systems are used to correct the phase.
The expression of the pupil of telescope $i$ apodized by the fibre mode in Eq.~\ref{Eq:ApodisedPupil} can be adjusted to reflect the effect of the turbulent phase $\phi_i$ as follows:
  \begin{equation}
    P_{\varocircle,i}(\vv{u}) = \Pi_\varocircle e^{i\phi_i(\vv{u})},
    \label{Eq:TurbulentApodisedPupil}
  \end{equation}
where $\Pi_\varocircle$ is the aberration-free pupil apodised by the fibre mode, which is assumed identical for all telescopes.
Injecting this definition in the expressions of the photometric lobe ($L(\vv\alpha)$, Eq.~\ref{Eq:PhotometricLobe}) and interferometric lobe ($L_{12}(\vv\alpha)$, Eq.~\ref{Eq:InterferometricLobe}), and taking a long-exposure average $\left< \right>_\mathrm{long}$ on the turbulent phase $\phi$, we obtain
  \begin{eqnarray}
    \left< L(\vv\alpha) \right>_\mathrm{long}
      & = & \invft{\left(\Pi_\varocircle \otimes \Pi_\varocircle\right)(\vv{u}) \; e^{-\frac{1}{2}D_\phi(\vv{u})}}, \\
    \left< L_{12}(\vv\alpha) \right>_\mathrm{long}
      & = & \invft{\left(\Pi_\varocircle \otimes \Pi_\varocircle\right)(\vv{u}) \; e^{-\sigma^2_\phi}},
      \label{Eq:Lobe_long}
  \end{eqnarray}
where $D_\phi(\vv{u})$ is the structure function of the turbulent phase as defined in \citet{Roddier1981}. The correlations of the phase within the telescope pupil, set by the phase structure function, are responsible for the widening of the single-mode photometric lobe.  We have made two strong assumptions to derive these results. First, the integration time is long enough to be considered infinite, meaning that the zero-mean hypothesis is correct. Second, we have assumed that the properties of turbulence are stationary, although they may vary with time in practice. The above derivation is valid only during the period over which this assumption is correct. These expressions can be upgraded for non-stationary turbulence by summing the lobes over finite timescales where turbulence can be assumed to be stationary. The underlying differences between the photometric and interferometric lobes will therefore remain.
When the baseline between the two apertures is larger than the outer scale of the turbulence, as is the case for GRAVITY on the Unit Telescopes, the phase in the two apertures becomes uncorrelated because the structure function $D_\phi(\vv{u})$ saturates to $2\sigma^2_\phi$.
Contrary to the single-mode photometric lobe, the angular size of the single-mode interferometric lobe does not change and only becomes attenuated by the coherent energy.
This behaviour is similar to the non single-mode case, as observed by \citet{Rousset+1992}, and was first demonstrated by \citet{Mege2002}.

As shown by \citet{Guyon2002} in the case of a point source, the visibility function of the apodised object can be recovered if the coherent energy can be accurately calibrated.
The loss of the signal-to-noise ratio on the visibility measurements increases exponentially with turbulence strength, however.
\citet{Guyon2002} also showed that the traditional calibration becomes difficult to impossible when the object size is comparable to or larger than the fibre mode.
An unresolved calibrator will not allow calibrating the widening of the photometric lobe.
It is then mandatory to measure the lobe in all the directions of the object with the same exposure time, assuming turbulence characteristics have not changed in the meantime.
This is discussed in Section~\ref{Sec:photometric_calibration}.

\section{Effects of partial phase correction}
\label{Sec:5}

\subsection{Effect on the interferometric field of view}

Single-mode interferometers are rarely used in fully turbulent conditions.
The phase variations across the telescope apertures are either partially corrected for by an adaptive optics system and/or modified by additional perturbations of instrumental origin.
As an illustration, we consider a scenario where the atmospheric turbulence is almost fully corrected for by an adaptive optics system, and the phase disturbance is dominated by turbulence, mostly tip-tilt, generated inside the optical train of the interferometer.
This is the case when a bright object is observed with VLTI (Very Large Telescope Interferometer)/GRAVITY \citep{1stlight}. Although the assumption that tunnel turbulence shares the behaviour of free air turbulence is quite strong, it is reasonable to assume that the statistics of tip-tilt residuals are stationary and average down to zero over relatively short timescales.

The phase in the pupil of telescope $i$ writes
  \begin{equation}
    \phi_i(\vv{u}) = 2\pi\vv{t}_i\cdot\vv{u} + p_i
    \label{Eq:22}
  ,\end{equation}
where $\vv{t}_i=(t_{i,x},t_{i,y})$ are the time-varying angular tip and tilt coefficients and $p_i$ is the time-varying piston term.
The instantaneous cross correlation of the apodised pupils function then writes
  \begin{equation}
    \mathrm{P_{\varocircle,1}} \otimes \mathrm{P_{\varocircle,2}} = \left[\mathrm{\Pi_\varocircle}e^{-i (2\pi\vv{t}_1\cdot\vv{u}+p_1)}\right] \otimes \left[\mathrm{\Pi_\varocircle}e^{-i (2\pi\vv{t}_2\cdot\vv{u}+p_2)}\right].
    \label{Eq:23}
  \end{equation}
The tip, tilt, and piston are assumed to be independent and to follow Gaussian distributions.
The standard deviations are $\sigma_\mathrm{t}$ for each tip-tilt axis and $\sigma_\mathrm{p}$ for the differential piston.
If the turbulence is decorrelated between the two pupils, the long-exposure average $\left< \right>_\mathrm{long}$ of the above expression writes
  \begin{equation}
    \left< P_{\varocircle,1} \otimes P_{\varocircle,2} \right>_\mathrm{long} =
      e^{-\frac{1}{2}\sigma_\mathrm{p}^{2}}
      \left[\Pi_\varocircle e^{-2(\pi u \sigma_\mathrm{t})^2} \right] \otimes
      \left[\Pi_\varocircle e^{-2(\pi u \sigma_\mathrm{t})^2} \right],
    \label{Eq:26}
  \end{equation}
where $u$ denotes the norm of $\vv{u}$.

The average tip-tilt effect is to multiply the apodised pupils by Gaussian functions of the spatial frequencies.
For a perfectly Gaussian fibre mode in image space,
  \begin{equation}
    E_\mathrm{fib}(\vv{\alpha})=e^{-\frac{\alpha^2}{2k^2}},
  \end{equation}
where $k$ is adjusted\footnote{For an optimal coupling between a single-mode Gaussian profile and a telescope, the size of the Gaussian beam $k$ has to match the telescope diameter $D$ following $k = \frac{\lambda}{D} \frac{\epsilon}{\sqrt{4\ln{2}}}$ , where $\lambda$ is the wavelength and $\epsilon$ a factor that depends on the shape of the pupil or on the central obscuration if the pupil is circular. The $\epsilon$ factor is close to unity, and the FWHM of the fibre mode almost perfectly matches the diffraction limit.} to maximise the coupling of astronomical light from an unresolved object to a single-mode fibre in the diffraction-limited case \citep{CoudeDuForesto+2000}, the long-exposure apodised pupil affected by tip-tilt writes:
  \begin{equation}
    \begin{split}
      \Pi_\mathrm{\varocircle,t}
        & = \Pi_\varocircle \; e^{-2(\pi u \sigma_\mathrm{t})^2} \\
        & = \Pi \; \invft{ E_\mathrm{fib} }^{\star} e^{-2(\pi u \sigma_\mathrm{t})^2} \\
        & = \Pi \; e^{-2(\pi u)^2 (k^2+\sigma_\mathrm{t}^2)} \\
    \end{split}
  .\end{equation}

 \begin{figure}
    \includegraphics[width=\linewidth]{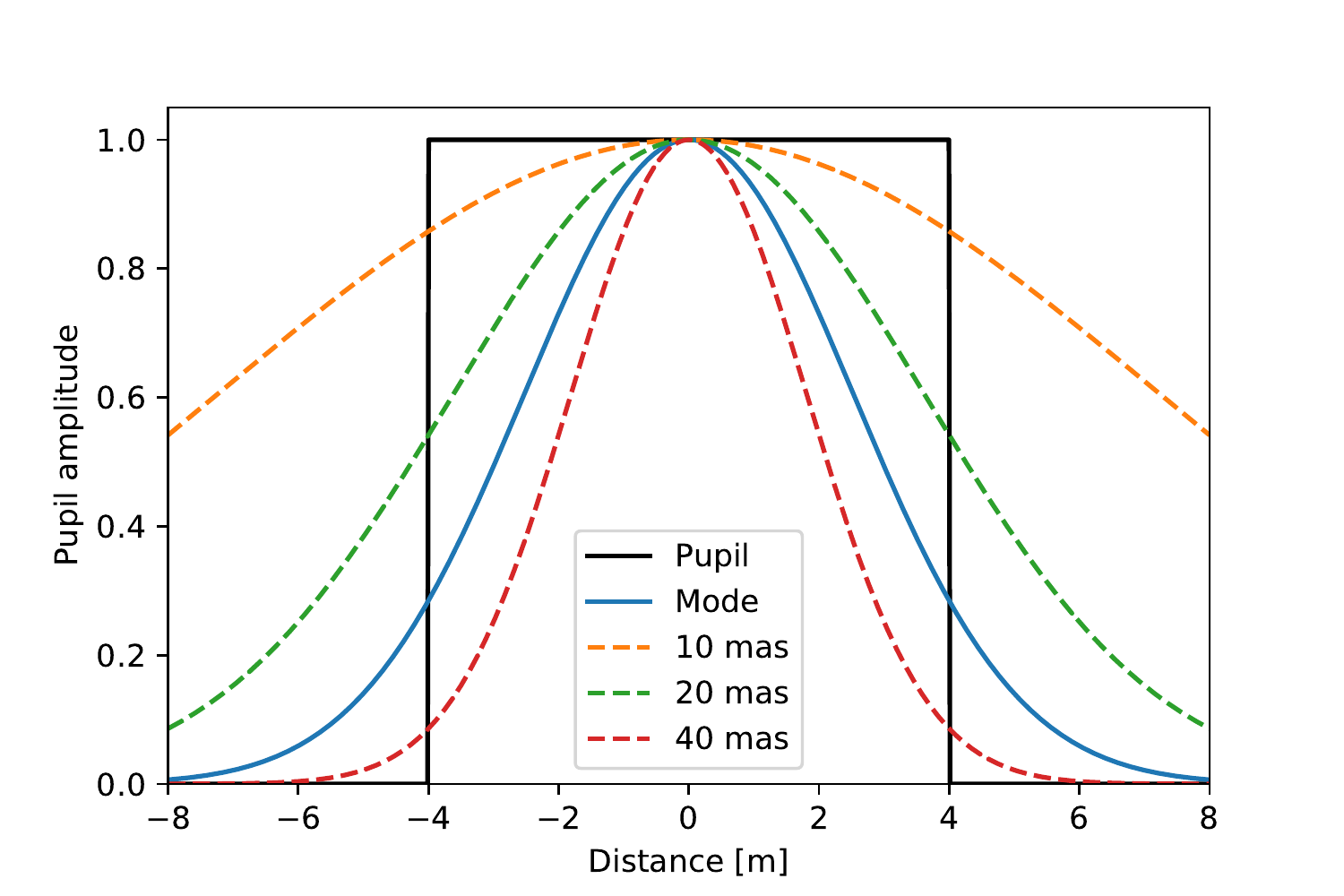}
    \includegraphics[width=\linewidth]{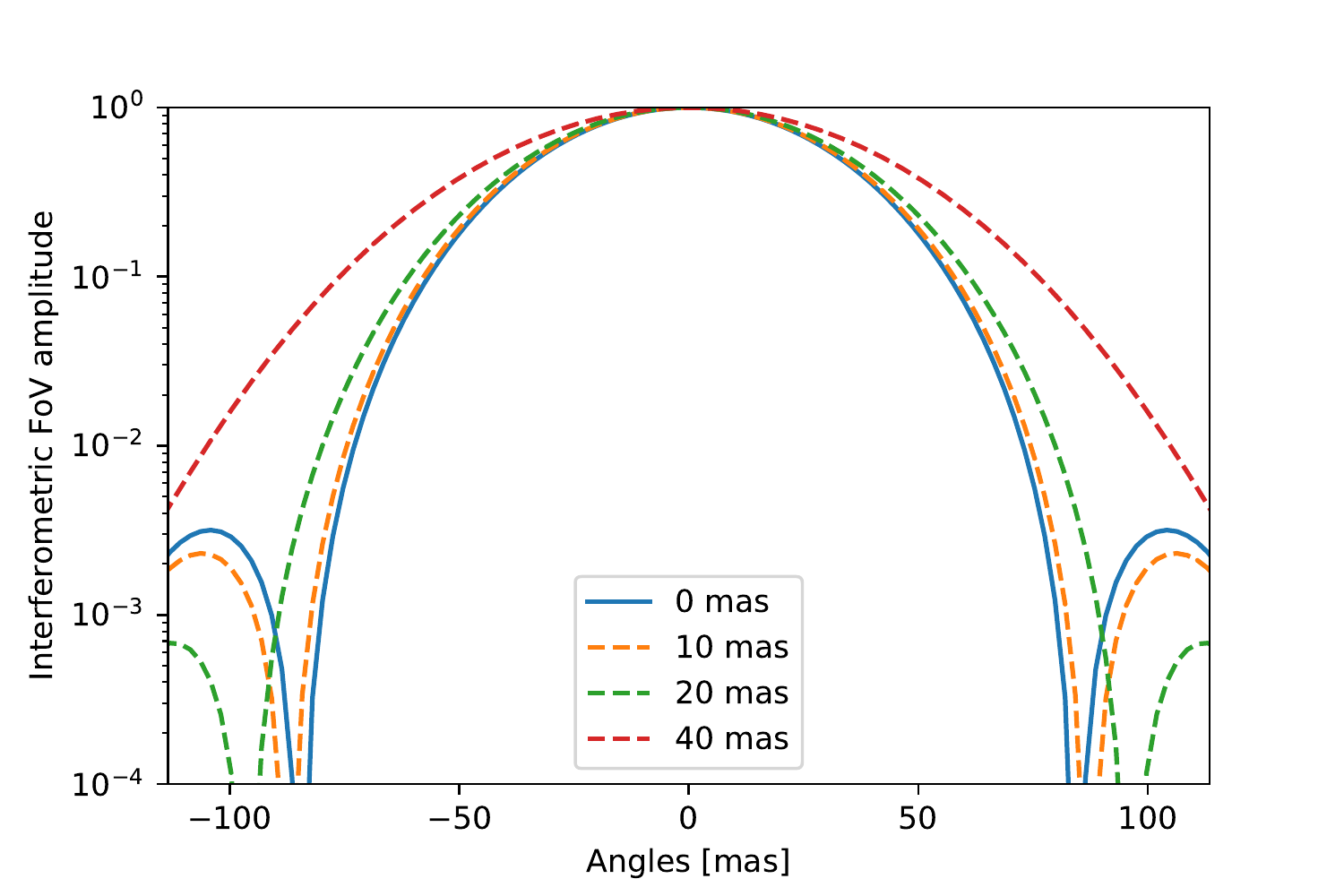}
    \includegraphics[width=\linewidth]{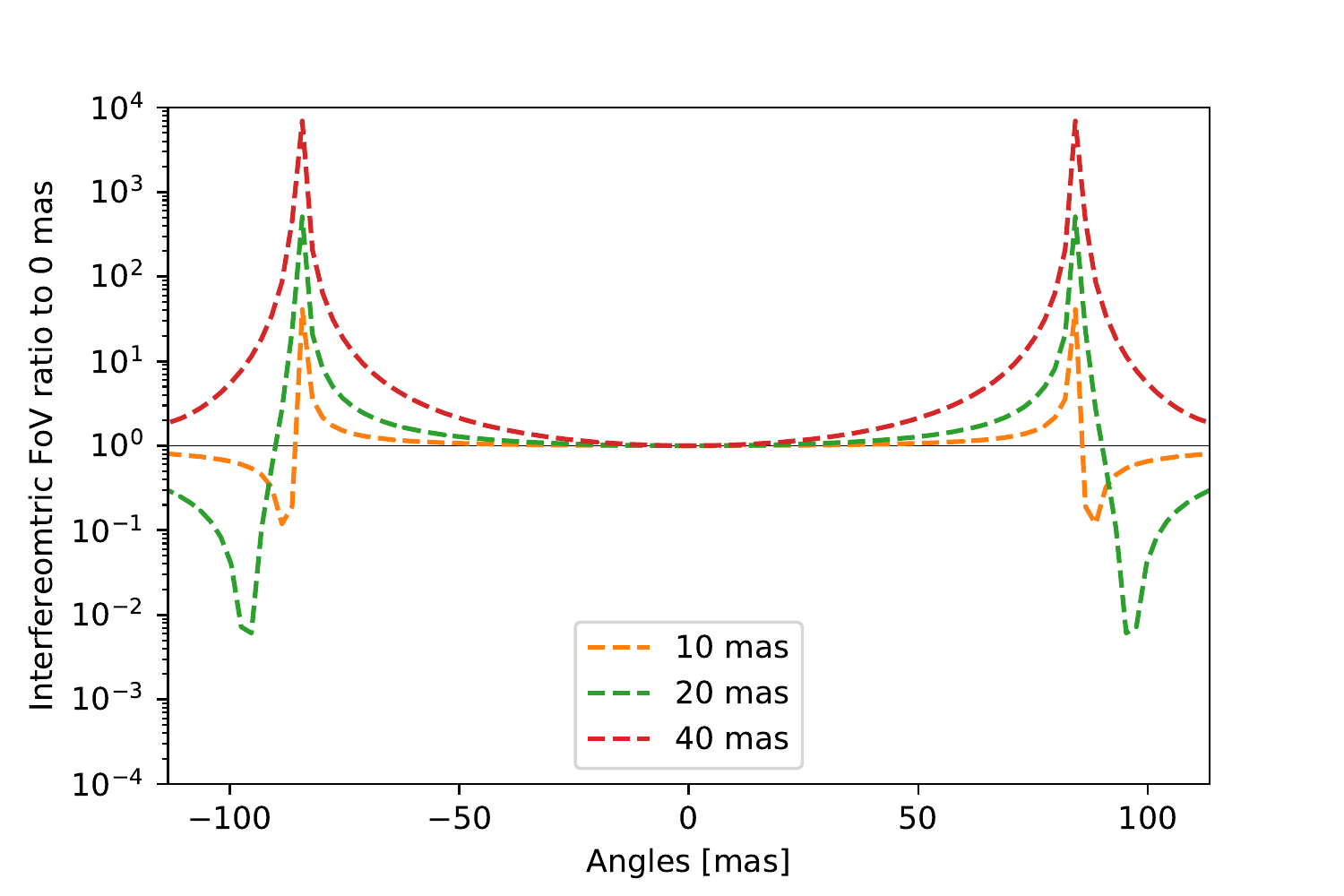}
    \caption{
      \textbf{Top}: Normalised profile in pupil plane of a Gaussian fibre mode (blue) that optimally matches a VLTI/UT-like $8\,\mathrm{m}$ telescope pupil (black) where the central obscuration has been neglected. The additional tip-tilt-induced apodisation profiles are given for 10, 20, and $40\,\mathrm{mas}$. The apodisation function is the product of the fibre mode and of the tip-tilt apodisation function.
      \textbf{Middle}: Interferometric lobes for $0\,\mathrm{mas}$, 10, 20, and $40\,\mathrm{mas}$ of tip-tilt, shown in logarithmic scale. Each lobe amplitude is normalised on axis and therefore does not illustrate the coupling performance reduction as the tip-tilt perturbation increases. As shown in the text, the photometric lobe matches the interferometric lobe when tip-tilt dominates.
      \textbf{Bottom}: Ratio between the 10, $20,$ and $40\,\mathrm{mas}$ tip-tilt lobes and the lobe without tip-tilt, shown in logarithmic scale. All tip-tilt rms values are for the total tip-tilt in all figures.
    }
    \label{Fig:TipTiltApodisation}
  \end{figure}

The effect of the residual tip-tilt is to further narrow the width of the apodised pupil function, whose Fourier transform is consequently enlarged.
A new interferometer lobe function can be defined in the same way as the fibre lobe was defined in Eq.~\ref{Eq:InterferometricLobe}:
  \begin{equation}
  \label{Eq:Interferometric_tip-tilt}
    L_\mathrm{12,t}(\vv{\alpha})
      = \invft{\Pi_\mathrm{\varocircle,t} \otimes \Pi_\mathrm{\varocircle,t}} (\vv\alpha)
    .\end{equation}
In absence of tip-tilt residues, the interferometer lobe matches the aberration-free fibre lobe. In the general case, however, its on-sky FWHM increases with the tip-tilt amplitude as follows:
  \begin{equation}
    \mathrm{FWHM}_\mathrm{t}
        = \sqrt{4\ln{2}\left(k^2+\sigma^2_\mathrm{t}\right)}
        = \sqrt{{ \left(\epsilon\frac{\lambda}{D} \right) }^2+4\ln{2}\sigma^2_\mathrm{t}}.
  \end{equation}

This result is new and has fundamental implications for the observation with single-mode interferometers of extended objects such as the Galactic centre.
Sources outside the diffraction limit of a single telescope but inside $\mathrm{FWHM_t}$ do contribute to the coherent flux.
Illustrations of this effect on the Unit Telescopes of the VLTI for different levels of residual tip-tilt are presented in Figure~\ref{Fig:TipTiltApodisation}.
We recall that the uncorrected tunnel tip-tilt at the VLTI is of the order of $\sim40$ $\mathrm{mas}$ per axis in the worst case, but is reduced to about $\sim10$ $\mathrm{mas}$ per axis when the tip-tilt is controlled with the acquisition camera of GRAVITY.

\begin{figure}
    \includegraphics[width=\linewidth]{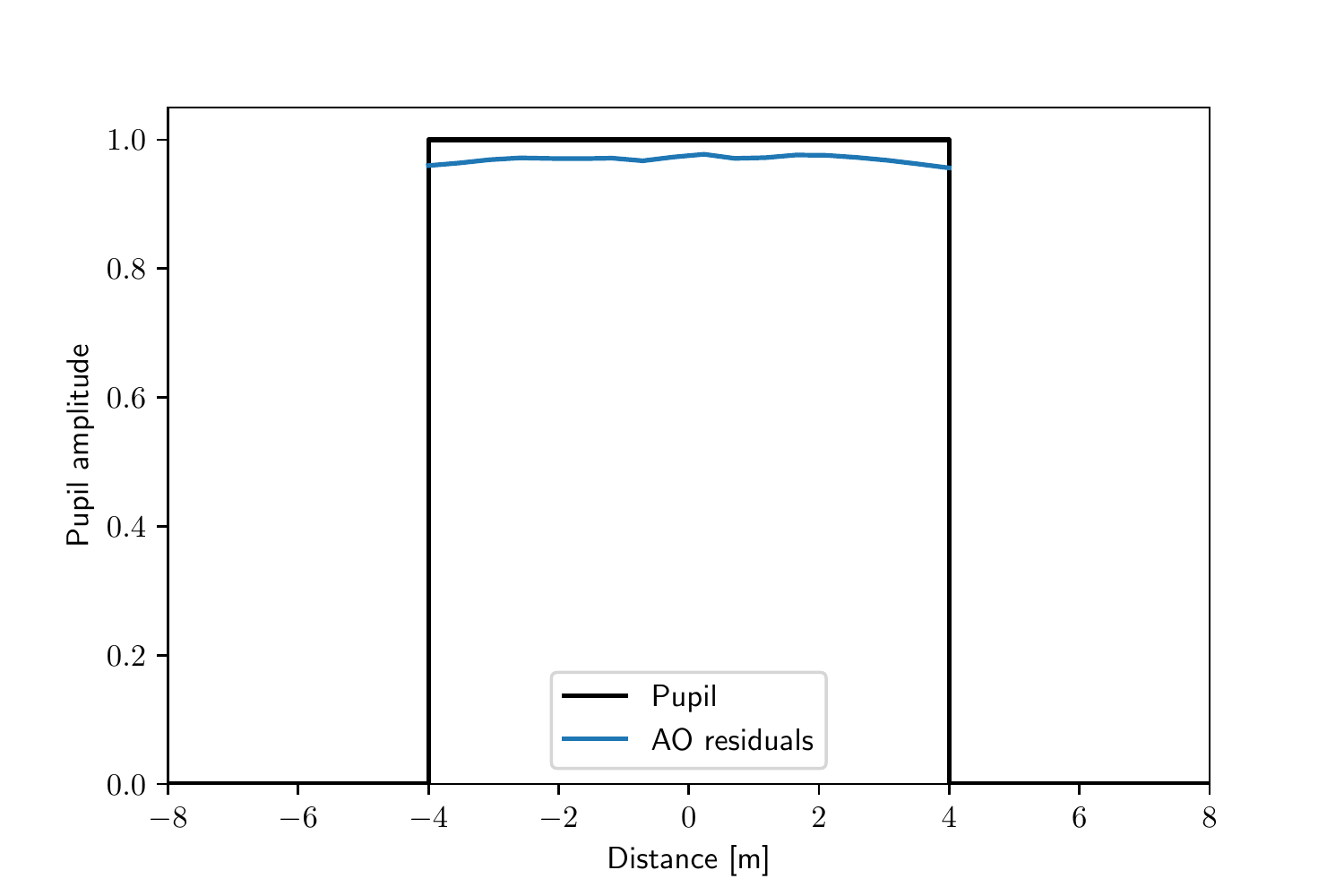}
    \caption{Pupil apodisation associated with the CIAO residual turbulence for a typical observation of the Galactic centre. There is an indication that the low orders (tip-tilt) are less strongly corrected for that the higher orders. Overall, the residual turbulence is not responsible for any significant widening of the interferometric field of view.}
    \label{Fig:CiaoApodisation}
\end{figure}

The phase variance of the modes of turbulent phase decreases with increasing order so that in practice, residual tip-tilt contains most of the phase variance after correction.
The influence of higher modes is therefore expected to be low at high Strehl ratio, as is the case with the CIAO (Coud\'e Infrared Adaptive Optics) adaptive optics system of GRAVITY.
To illustrate this, we now consider, in addition, the residual turbulent phase measured by the infrared wavefront sensor CIAO when the loop on IRS7 is closed in the Galactic centre.
In the particular case of CIAO and GRAVITY, Figure~\ref{Fig:CiaoApodisation} shows that the effect of modes higher than tip-tilt is rather negligible on the apodisation of the pupil and therefore on the interferometric field of view.

\begin{figure*}
    \centering
    \includegraphics[width=0.45\linewidth]{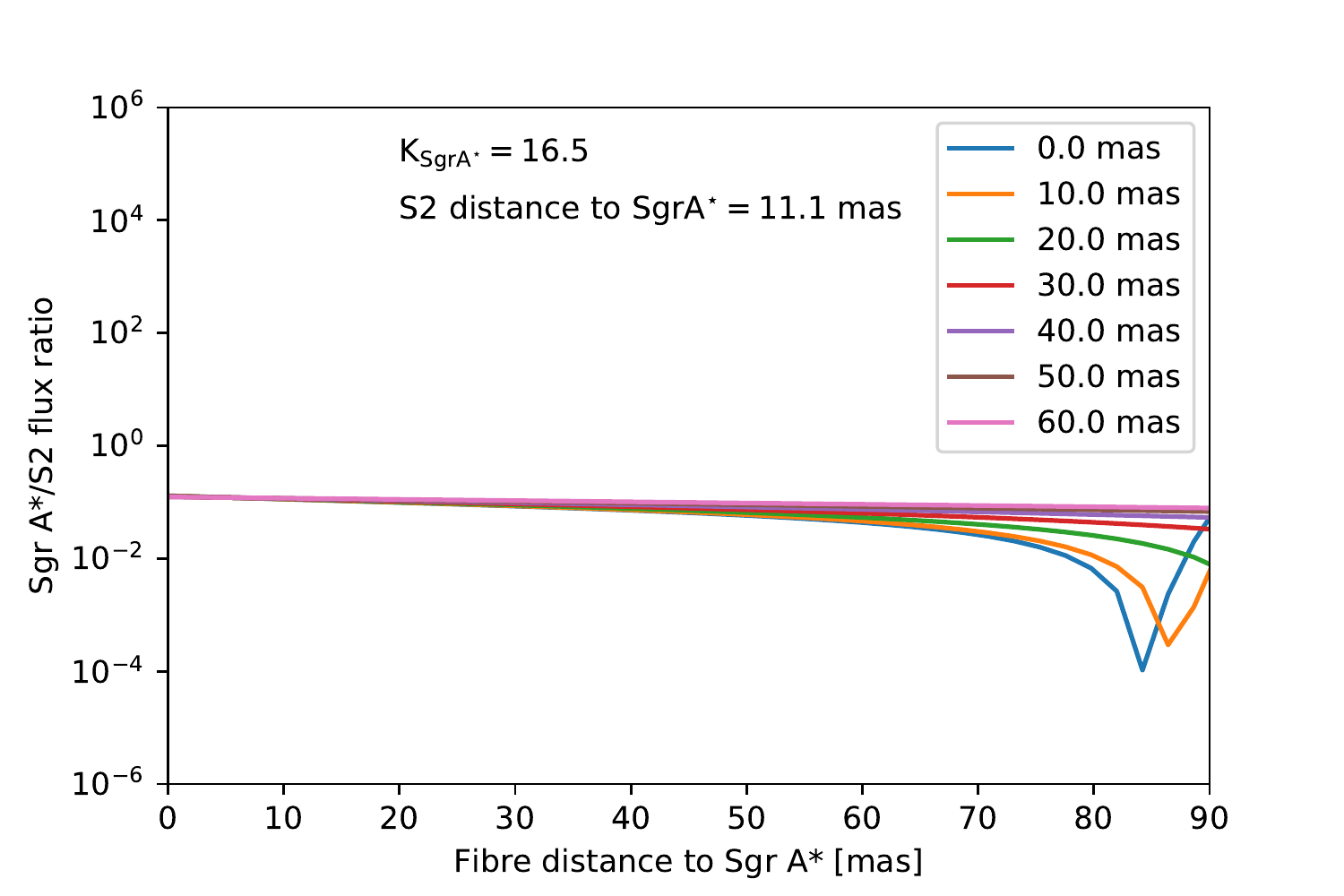}
    \includegraphics[width=0.45\linewidth]{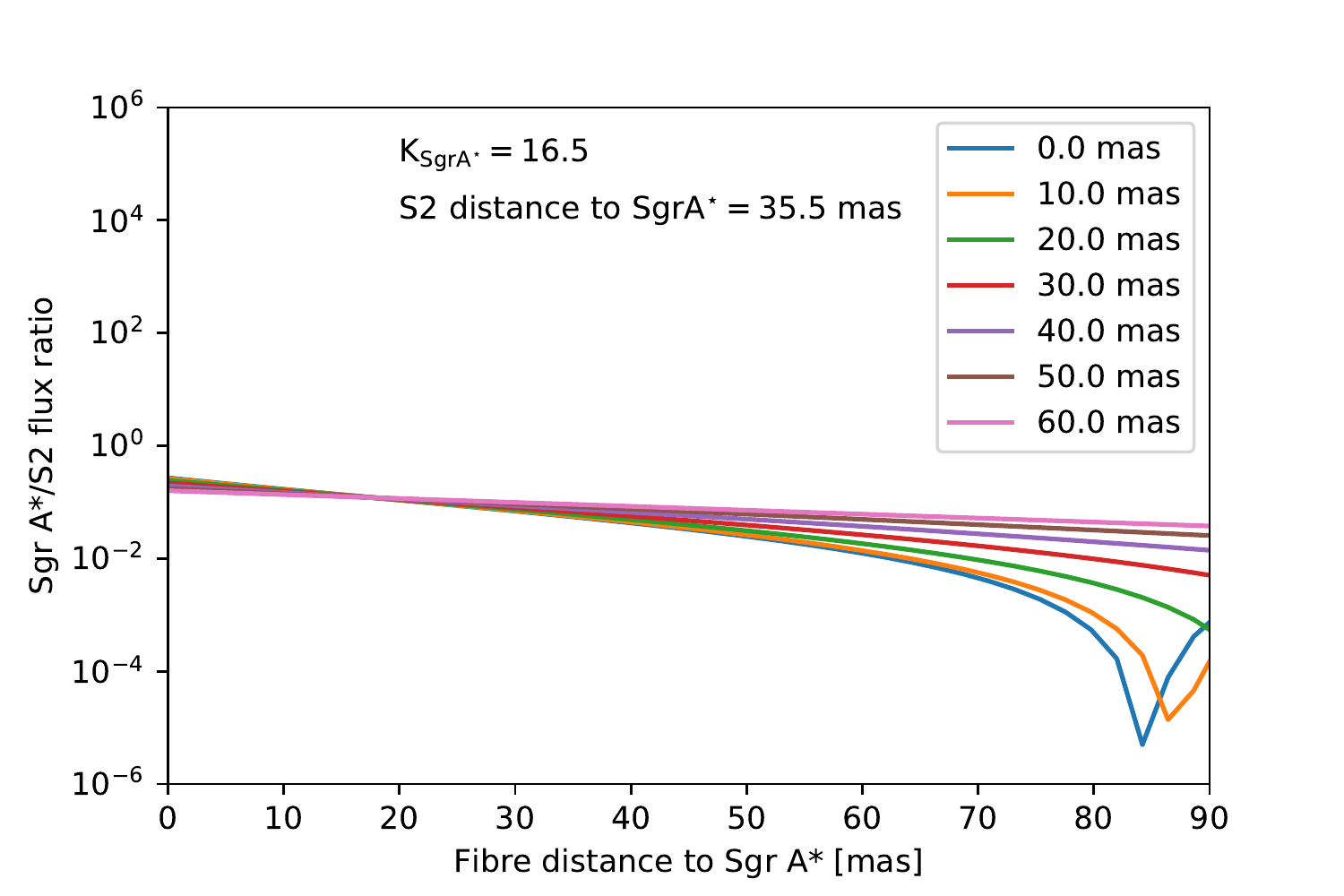}
    \includegraphics[width=0.45\linewidth]{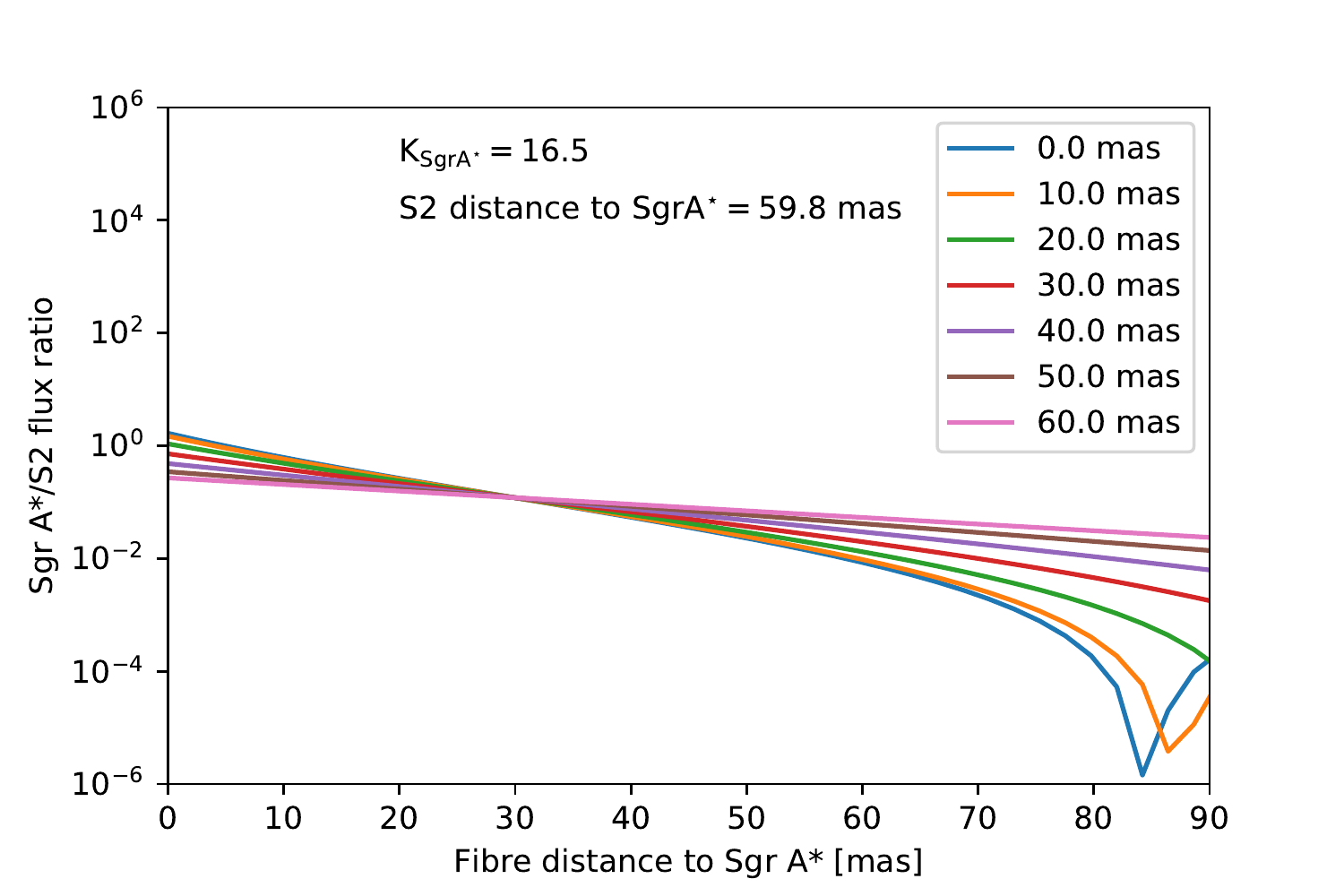}
    \includegraphics[width=0.45\linewidth]{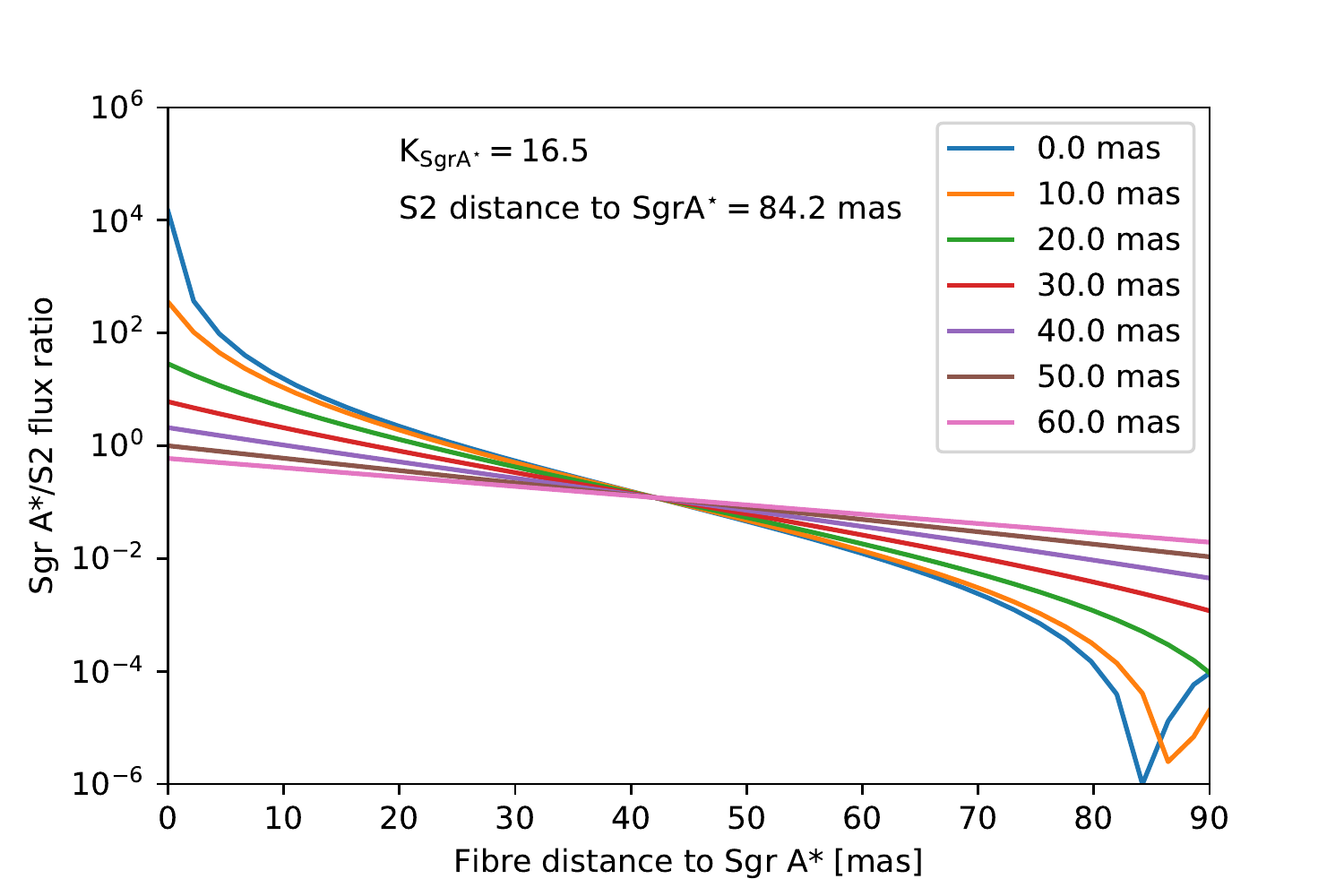}
    \caption{Field-of-view effect: widening by tip-tilt residues on the contrast ratio between S2 and a quiescent Sgr A$^\star$ as a function of relative distance and tip-tilt residue rms.}
    \label{Fig:S2_SgrA*_ratio_quiet}
\end{figure*}

The tip-tilt effect can be intuitively explained.
The high-order turbulence modes do not affect the image position to zeroth order but destroy the spatial coherence of light, with the consequence that speckles of various phases are introduced in the image.
The tip-tilt has a different effect.
It does not affect the instantaneous spatial coherence and only shifts the object position in the image plane.
In the two-telescope case, when the two images overlap, or equivalently, when the differential tip-tilt is of the order of the diffraction limit at most, then spatial coherence is high.
For a binary object where one of the components is at the centre of the turbulence-free field of view, the two objects are  equally affected by tip-tilt, but the effect on spatial coherence is different.
At the centre of the field of view, the lobe transmission is flat and maximum, and therefore the tip-tilt effect is that spatial coherence is reduced.
At the edge of the lobe, the throughput variation is steeper, and any positive tip-tilt fluctuation  (i.e. in the field centre direction simultaneously in both arms of the interferometer to ensure a good overlap of the images) leads to a strong increase in coherence throughput.

\subsection{Tip-tilt effect on the field of view around Sgr A$^{\star}$}

The net effect of widening the field of view by tip-tilt from tunnel seeing on the photometric ratio of Sgr A$^{\star}$ and S2 in the case of GRAVITY is represented in Figure~\ref{Fig:S2_SgrA*_ratio_quiet}.
We have used K magnitudes of 14.2 for S2 and 16.5 for Sgr A* in a quiescent state \citep{Genzel2003}, whereas the K magnitude reached 15.5 for the first moderate flare detected by GRAVITY \citep{1stlight} and can be as high as 14.5 for a brighter flare \citep{Genzel2003}.
The angular distance of S2 was set to various values ranging from $11\,\mathrm{mas}$, the closest distance at pericentre is $15\,\mathrm{mas}$ according to \citet{Gillessen2009}, up to $84\,\mathrm{mas}$ at the distance at the time of the observation reported in \citep{1stlight}.

The Sgr~A$^{\star}$-to-S2 flux ratio is plotted for various tip-tilt rms values and as a function of fibre offset from Sgr A$^{\star}$ in the direction of S2.
The rms tip-tilt measured by GRAVITY is typically 15\,mas under good seeing conditions, meaning that it is limited by tunnel seeing.
The plots show the net tip-tilt effect: it increases the relative contributions of the sources in the field of view around the central source.
For a separation of 84\,mas, the Sgr A$^{\star}$-to-S2 flux ratio is roughly $10^{4}$ in absence of tip-tilt and with the fibre centred on Sgr A$^{\star}$.
This same flux ratio drops to $10^{2}$ for a 15\,mas tip-tilt and reaches 1 when a fibre offset of 20\,mas is additionally applied.
Four orders of magnitude can therefore be compensated for with moderate tip-tilt and fibre offset. 
For shorter separations, the Sgr A$^{\star}$-to-S2 flux ratio is either systematically of the order of 1 or lower, with moderate rms tip-tilt applied.

Although the widening of the field of view found in Section~\ref{Sec:5} is small, the effect is quite large when the source at the centre of the field is much fainter than the source at the edge.
This clearly explains why S2 was detected and appeared so bright in the image reconstructed in \citet{1stlight}, despite the $85\,\mathrm{mas}$ offset.

\subsection{Strategies to widen the field of view}
\begin{figure}
    \includegraphics[width=\linewidth]{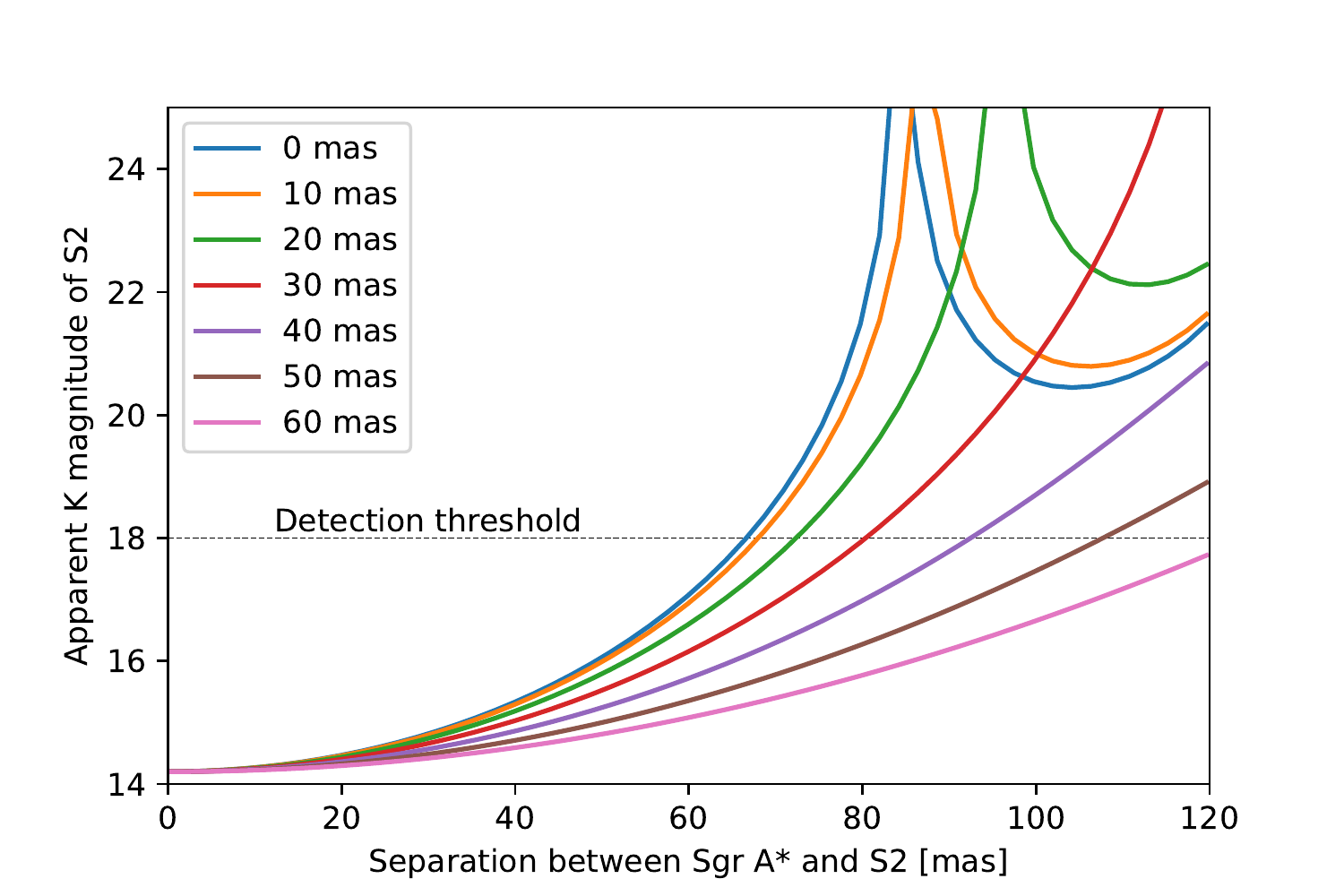}
    \caption{Tip-tilt effect on the apparent magnitude of S2 in the images reconstructed by GRAVITY centred on the Galactic centre. A 50\,mas tip-tilt allows keeping the apparent magnitude of S2 above the current sensitivity threshold of K=18 in the images reconstructed from GRAVITY data, up to a separation of $\sim$100\,mas.}
    \label{Fig:Apparent_K_S2}
  \end{figure}

\begin{figure}
    \includegraphics[width=\linewidth]{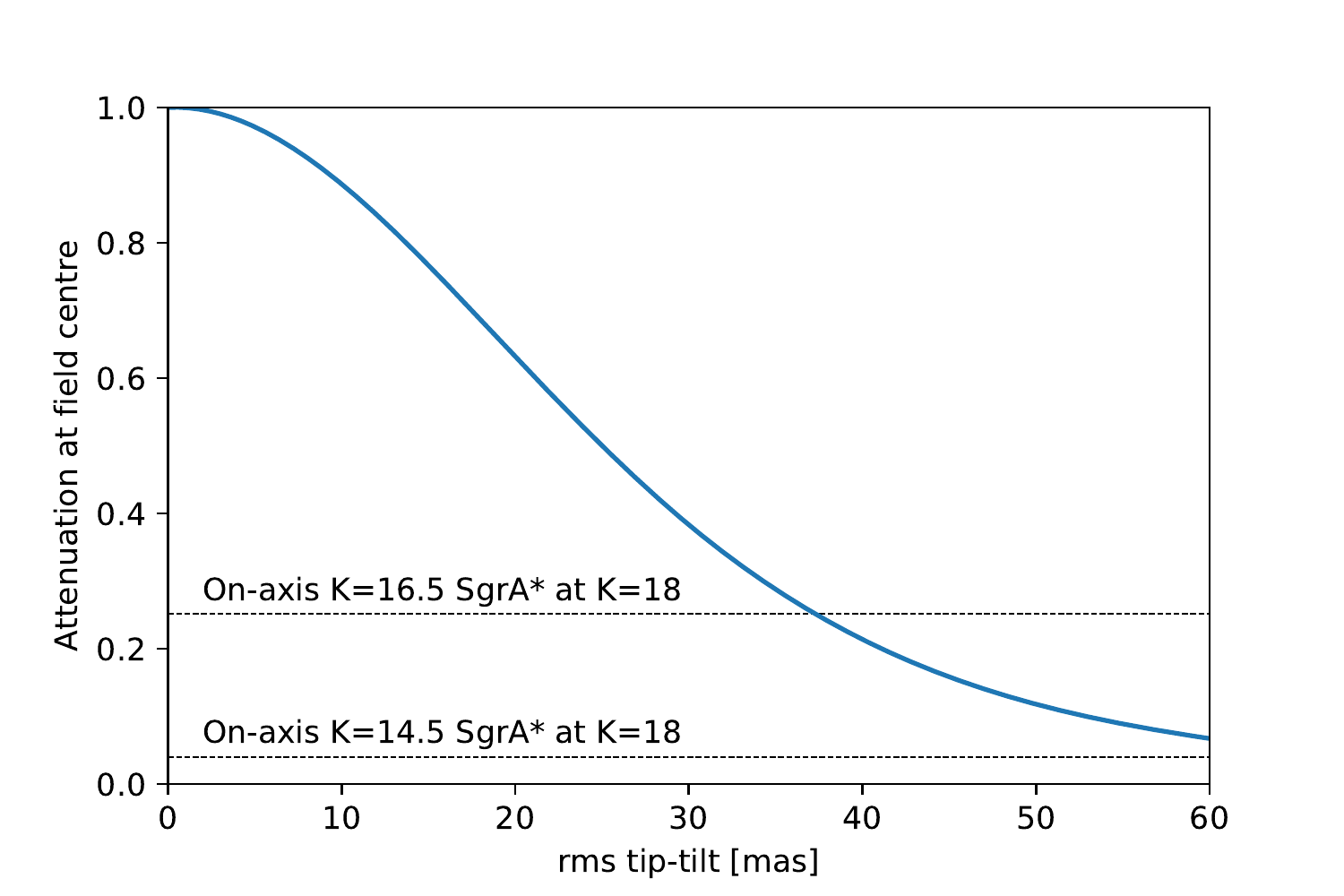}
    \caption{Tip-tilt effect on the attenuation of the interferometric throughput at the centre of the field of view. The figure also shows the level of attenuation where a quiescent K=16.5 or bright flaring K=14.5 Sgr A$^\star$ reaches the assumed K=18 detection limit. A quiescent Sgr A$^\star$ would no longer be observable above $\sim$37\,mas tip-tilt rms, whereas a bright flare would still be observable past 60\,mas.}
    \label{Fig:Attenuation_SgrA*}
  \end{figure}

\begin{figure*}
    \centering
    \includegraphics[width=0.45\linewidth]{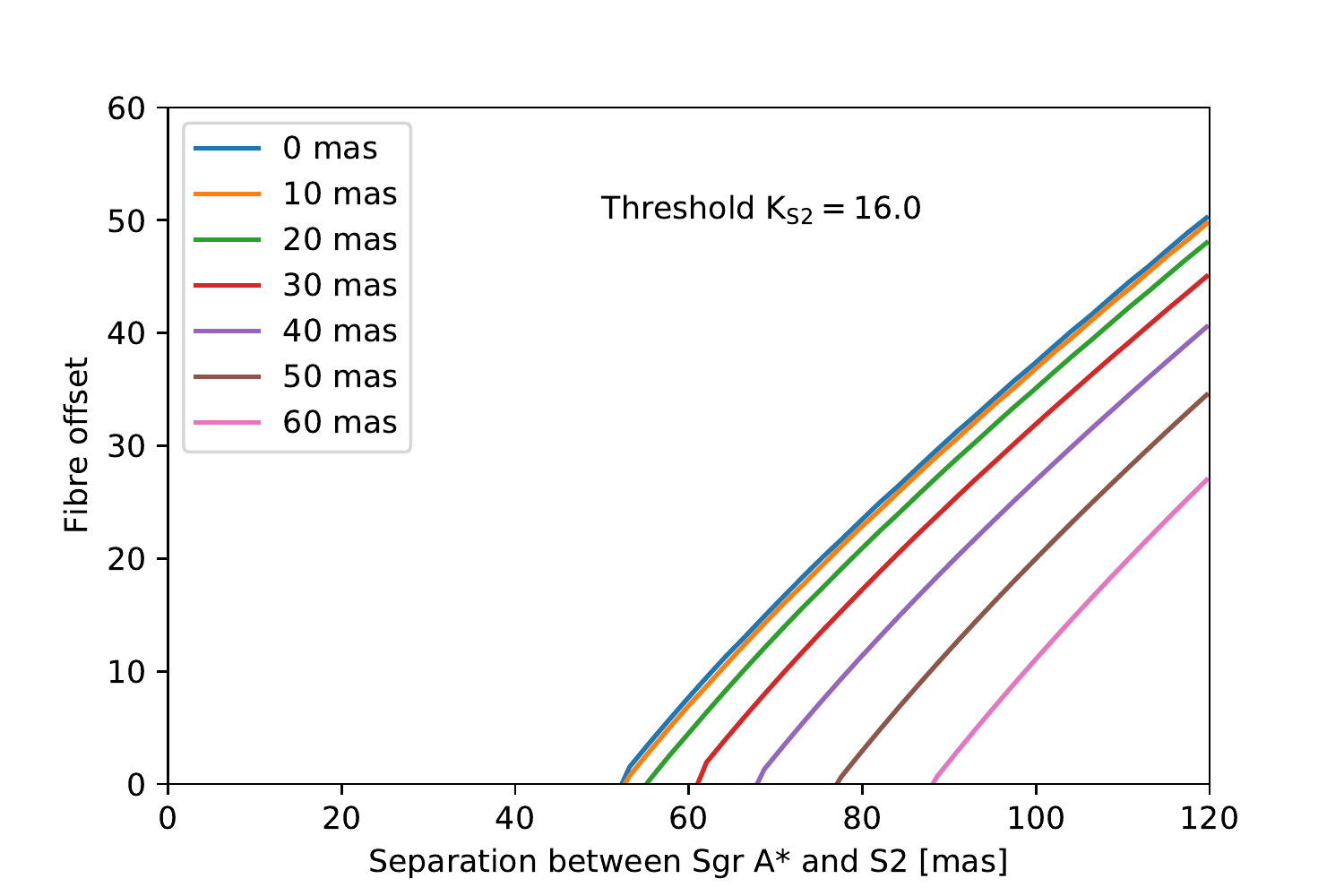}
    \includegraphics[width=0.45\linewidth]{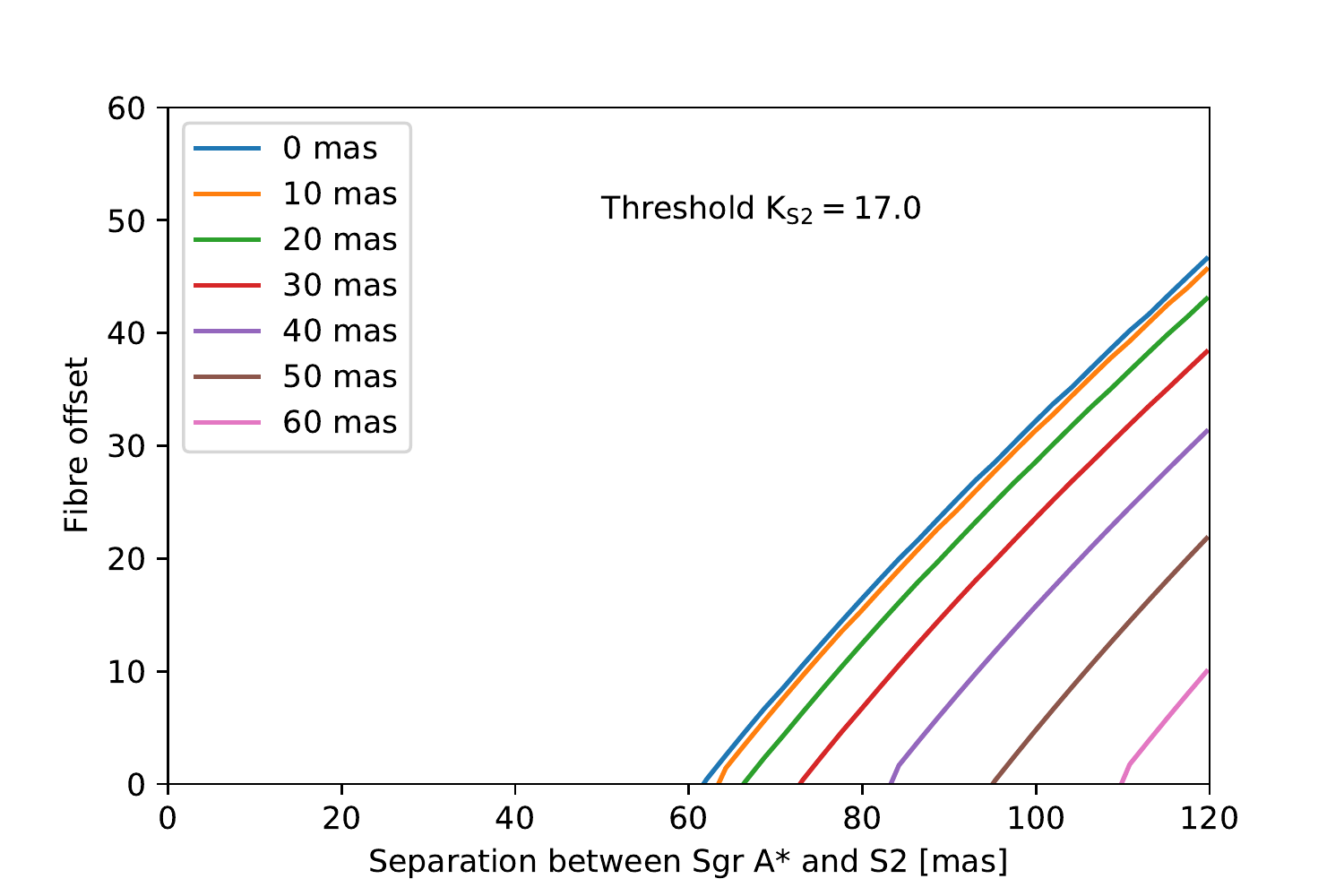}
    \includegraphics[width=0.45\linewidth]{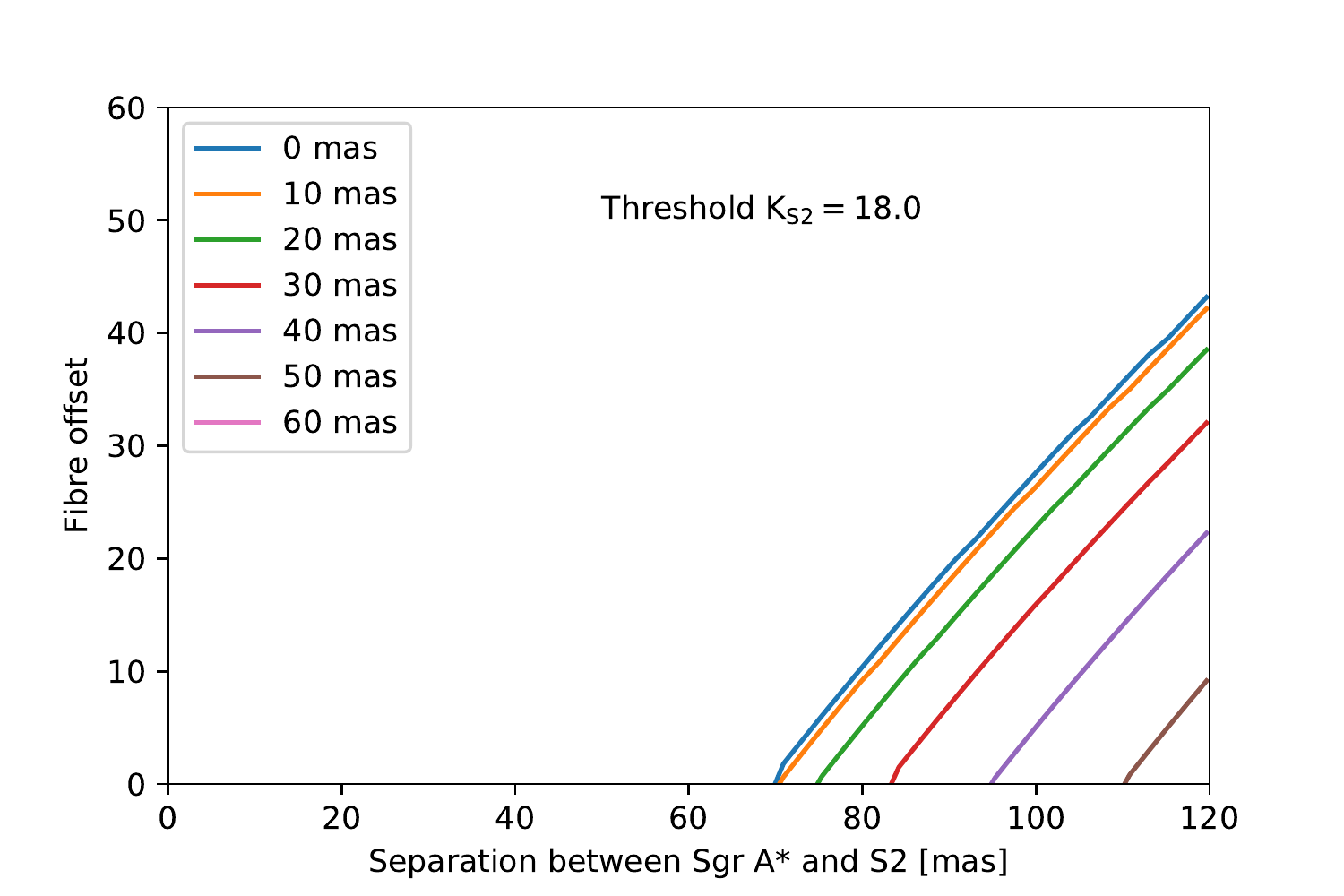}
    \includegraphics[width=0.45\linewidth]{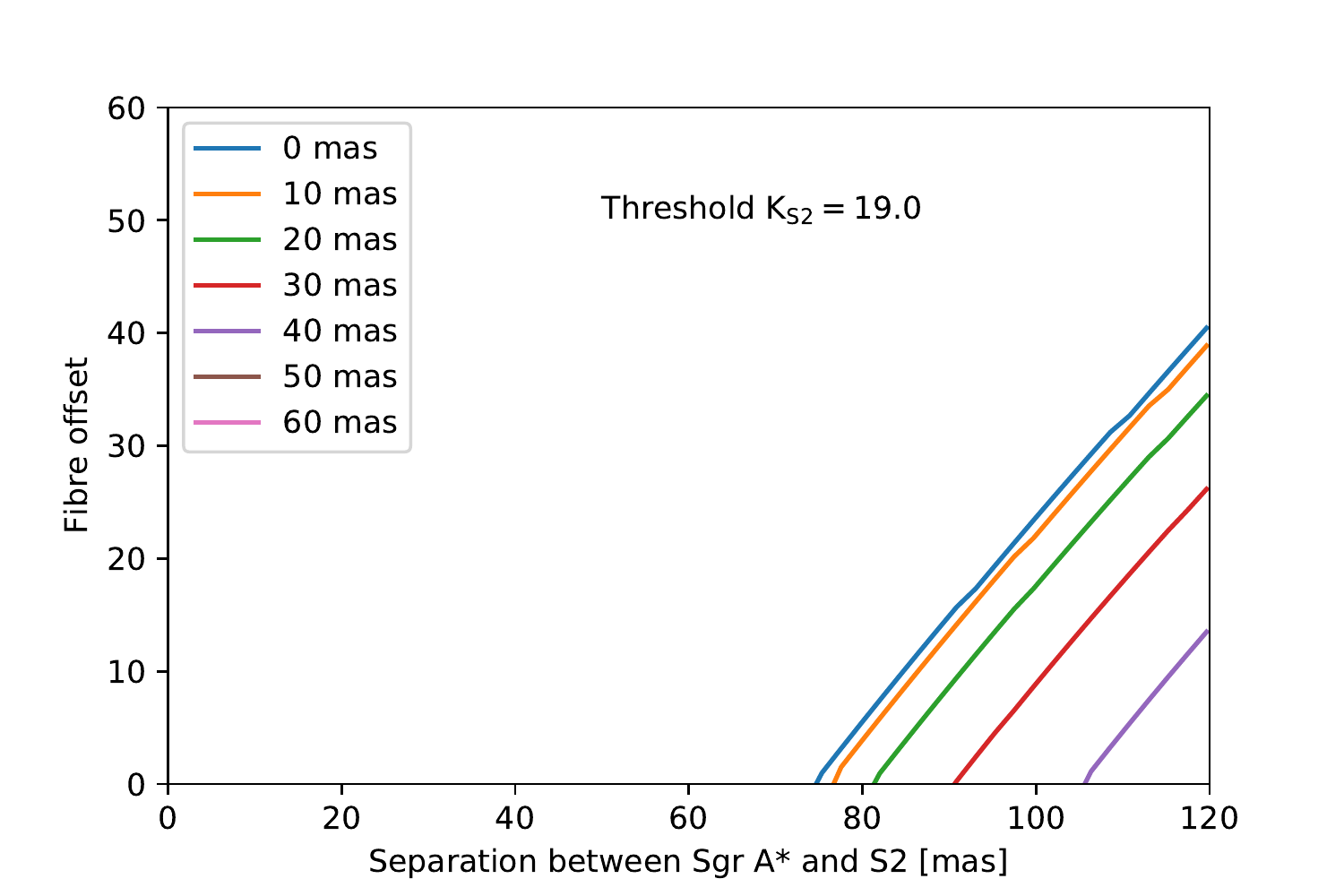}
    \caption{Fibre offset needed to reach the threshold magnitude to detect S2 in the reconstructed GRAVITY images as a function of rms tip-tilt. Results are plotted for four different threshold magnitudes: 16, 17, 18, and 19.}
    \label{Fig:Fiber_offset_S2}
  \end{figure*}

Another way to look at the tip-tilt effect on an interferometric field of view in the case of the Galactic centre is to study the apparent magnitude of S2, that is, the magnitude of S2 attenuated by the interferometric lobe function as it appears in a reconstructed image.
The current sensitivity of the reconstructed images of GRAVITY is K=17.5 \citep{1stlight}, and it will probably increase in the future.
Figure~\ref{Fig:Apparent_K_S2} gives the apparent magnitude of S2 as a function of rms tip-tilt and distance to Sgr A$^{\star}$, assuming the fibre is centred on Sgr A$^{\star}$.
Setting a threshold of K=18 for S2, we can read the rms tip-tilt required to detect S2 from this figure.
It increases with the separation between S2 and Sgr A$^{\star}$.
50\,mas of rms tip-tilt is required to detect S2 for a distance of 100\,mas.
This indeed provides a method to widen the field of view of the interferometer: apply random tip-tilt during exposures.
However, the effect of bandwidth smearing needs to be taken into account in this case because Sgr A$^{\star}$ and S2 are observed in the most sensitive mode of GRAVITY with the lowest spectral resolution $R=22$ with an intrinsic FWHM of the interferometric field of view of order 60\,mas ($R\times\frac{\lambda}{B}$).
The widening of the field of view through{\it } tip-tilt also affects the central source, as shown in Figure~\ref{Fig:Attenuation_SgrA*}.
The central source is all the dimmer as the tip-tilt rms is higher.
When in a quiescent state, Sgr A$^{\star}$ becomes as low as the threshold of K=18 for an rms tip-tilt of 50\,mas.
All in all, given the current sensitivity of the reconstructed images of GRAVITY, the field of view can be extended up to 120\,mas with an artificial tip-tilt of 37\,mas.
In addition to tip-tilt, the apparent magnitude of S2 in the reconstructed images can be increased by moving the fibre towards S2.
Figure~\ref{Fig:Fiber_offset_S2} shows the fibre offset that is to be applied to reach certain threshold magnitudes for S2 as a function of rms tip-tilt.
The offsets were computed for four different threshold magnitudes: 16, 17, 18, and 19.
The higher the threshold (and therefore the sensitivity of the GRAVITY reconstructed maps), the lower the required offset.
For a threshold of K=18, no fibre offset is required up to a separation between S2 and Sgr A$^{\star}$ of 70\,mas.
Above this separation, the offset required to detect S2 increases with decreasing rms tip-tilt. Another similar method is to perform a 2D scan of the field of view in a single exposure. This will be discussed in a forthcoming paper. This is an alternative to the well-known mosaicing method of radio astronomy.

\subsection{Effect on photometric calibration}
\label{Sec:photometric_calibration}

\begin{figure}
    \includegraphics[width=\linewidth]{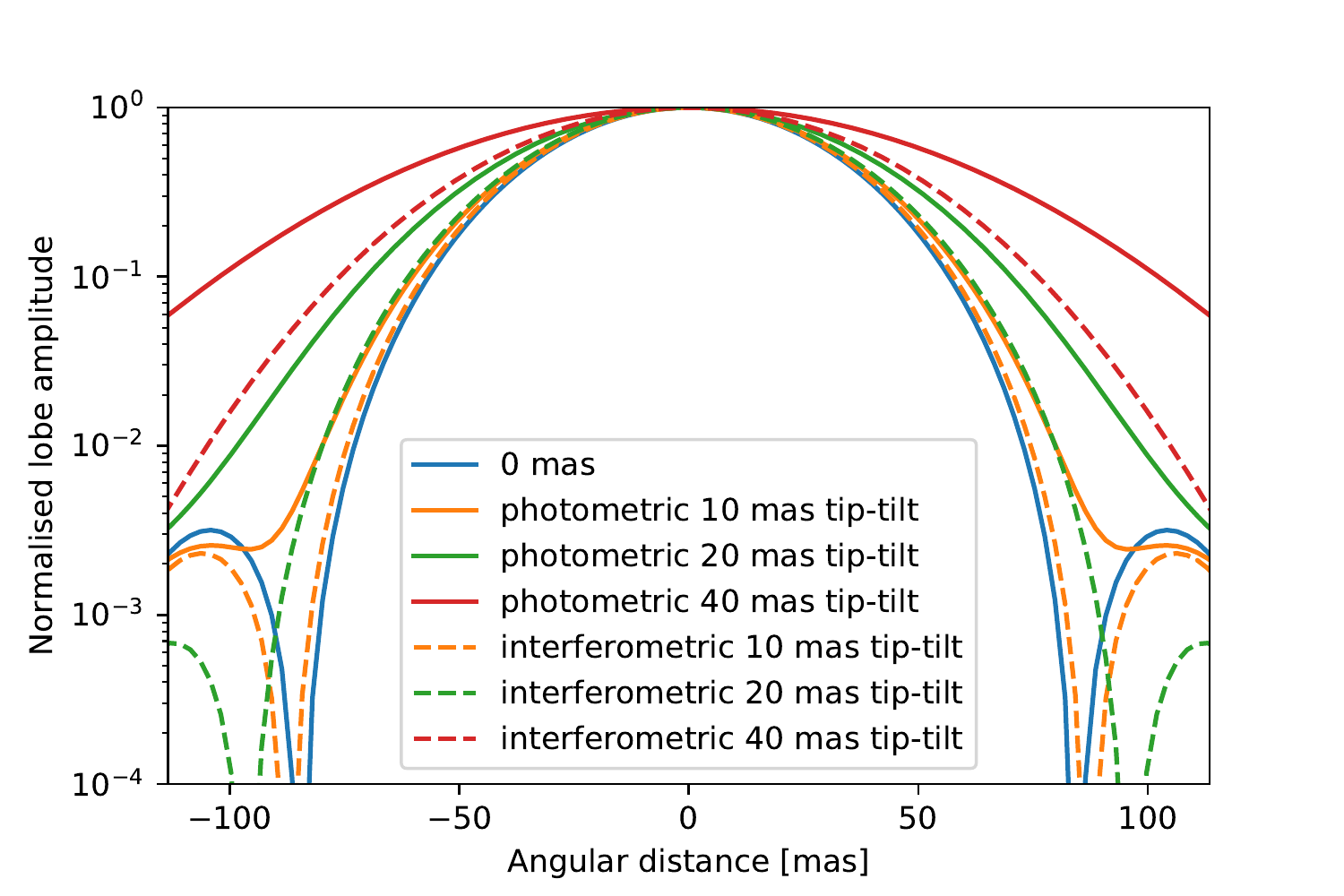}
    \caption{Photometric lobes for $0\,\mathrm{mas}$, $20\,\mathrm{mas}$, and $40\,\mathrm{mas}$ of tip-tilt, shown in logarithmic scale. Each lobe amplitude is normalised on axis and therefore does not illustrate the coupling performance reduction as the tip-tilt perturbation increases. The interferometric lobes of Figure~\ref{Fig:TipTiltApodisation} are plotted as dashed lines.}
    \label{Fig:TipTiltApodisation_phot}
  \end{figure}

A similar computation as in the previous section is applied to the photometric lobe when turbulence is dominated by tip-tilt alone.
In this case, the structure function of the turbulent phase takes the simple expression
    \begin{equation}
        D_{t}(\vv{u})=\left( 2 \pi \sigma_{t} u \right)^2
    ,\end{equation}
with the long-exposure photometric mode of Eq.~\ref{Eq:Lobe_long} becoming
    \begin{equation}
        \left< L_\mathrm{t}(\vv\alpha) \right>_\mathrm{long}=\invft{\left(\Pi_\varocircle \otimes \Pi_\varocircle \right)(\vv{u}) \; e^{-2\left( \pi \sigma_{t} u \right)^2}}
    .\end{equation}
As a consequence, the long-exposure photometric mode does not match the long-exposure interferometric mode of Eq.~\ref{Eq:Interferometric_tip-tilt}.
This is confirmed by the plots in Figure~\ref{Fig:TipTiltApodisation_phot}.
The reason for this is the non-correlation of the tip-tilt between the two interferometric pupils.
The effect is less dramatic than in Eq.~\ref{Eq:Lobe_long} for the case of uncorrected turbulence, however, where the interferometric lobe shape is independent of turbulent while the photometric lobe increasingly widens with increasing turbulence strength.
When the FWHM is computed for the 10\,mas rms tip-tilt case, the difference is not so great, however (70\,mas versus 67\,mas; 65\,mas in both cases without turbulence).

As pointed out by \citet{Guyon2002}, this difference between the interferometric and photometric lobes can make the photometric calibration of the single-mode visibilities difficult.
In the case of GRAVITY, with the low rms of residual tip-tilt and if the pointing of the fibre is accurate enough, the difference in lobe FWHM is so low that the effect could be negligible.
However, if the contrast between the central source and the sources at the edge of the field of view is high (as is the case if Sgr A$^{\star}$ is at the centre of the field and in quiescent or low-state mode), then the photometric correction of the visibilities can lead to a substantial visibility bias.
More photons are coupled to the photometric lobes than to the interferometric lobes, meaning that the number of coherent photons is overestimated.

\begin{figure}[H]
    \includegraphics[width=\linewidth]{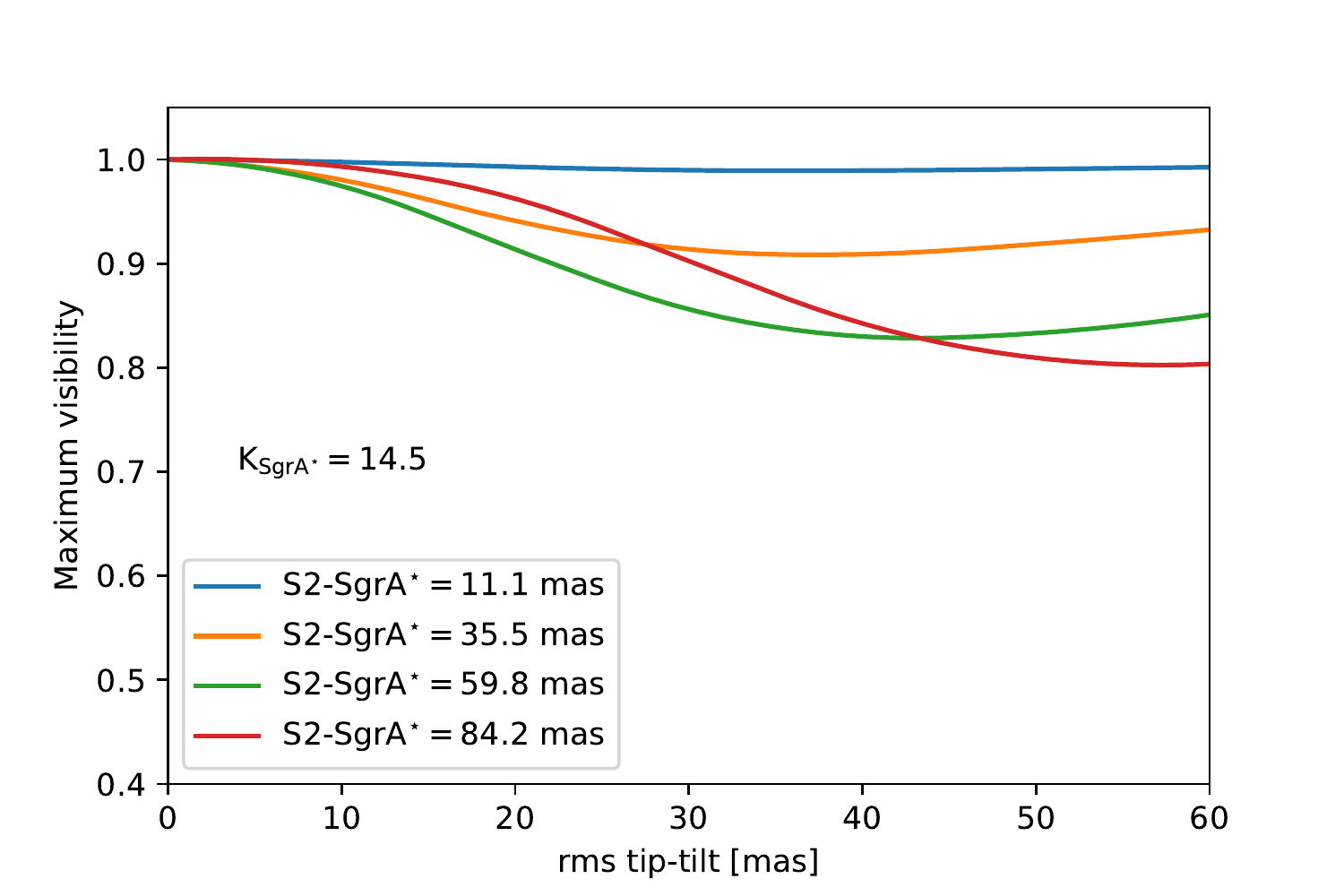}
    \includegraphics[width=\linewidth]{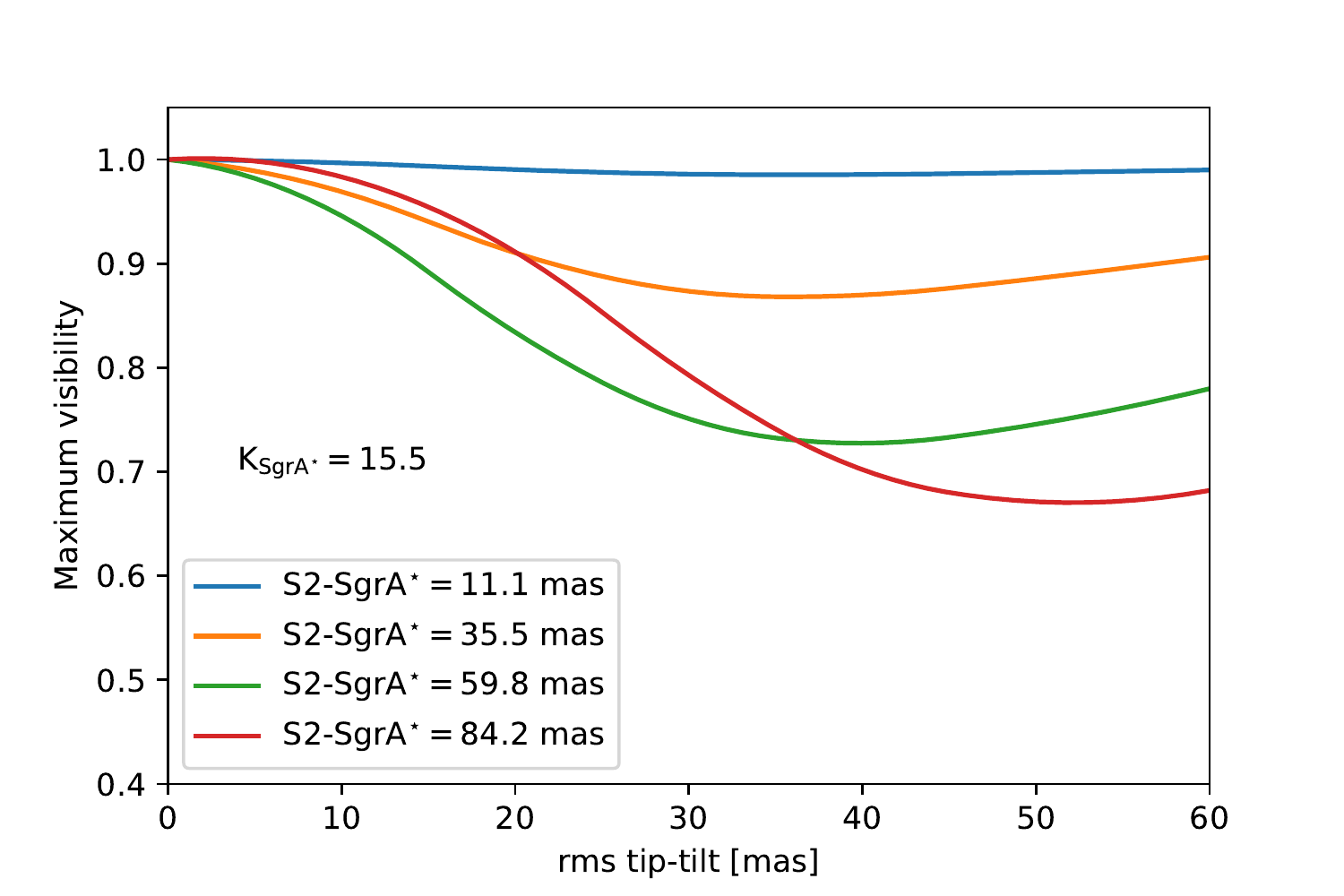}
    \includegraphics[width=\linewidth]{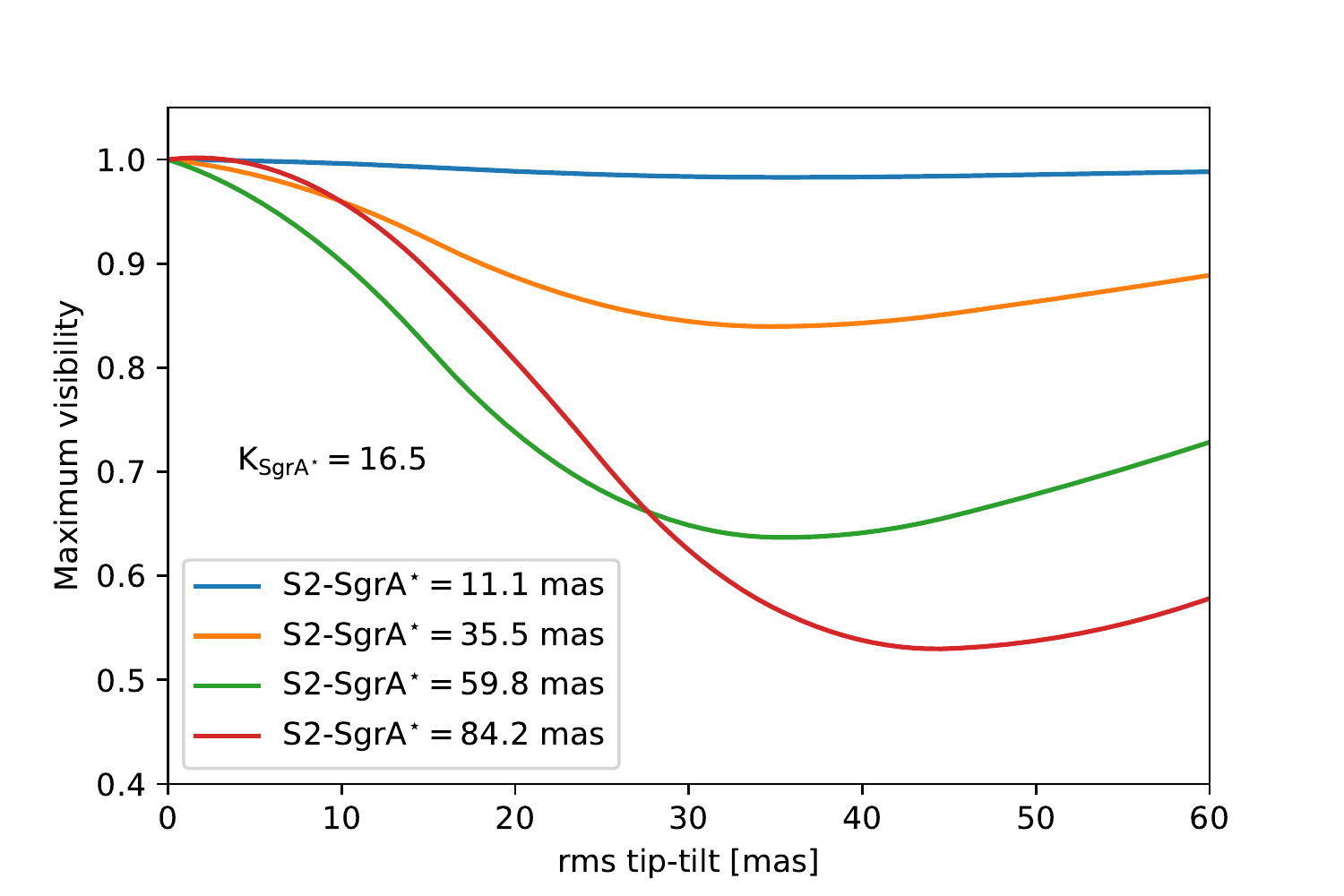}
    \caption{Effect of differential interferometric vs. photometric field-of-view widening caused by residual tip-tilt. The figure shows the maximum visibility measured for the S2 / Sgr A$^\star$ pair as a function of tip-tilt perturbation for four different separations between the two sources and for the three different states of Sgr A$^\star$ (top: bright flare; middle: moderate flare; bottom: quiescent state).}
    \label{Fig:max_contrast_quiet}
\end{figure}

This can be shown by again expanding the two-telescope balanced interferogram of Eq.~\ref{Eq:Interferogram}, but keeping the photometric terms:
    \begin{equation}
        I_{12} = I_1 + I_2 + 2 \Re \left( \gamma_{12} \right).
        \label{Eq:Interferogram2}
    \end{equation}
Injecting the photometric and interferometric lobes of Eq.~\ref{Eq:Photometry-Lobe} and Eq.~\ref{Eq:Cohrence-Lobe}, respectively, results in    \begin{equation}
        I_{12} = \ft{L_{1}O}(\vv{0}) + \ft{L_{1}O}(\vv{0}) + 2 \Re \left( \ft{L_{12}O}(\vv{B}/\lambda)  \right)
        \label{Eq:Interferogram3}
    .\end{equation}
Applying the Zernike-Van Cittert to the third term, we obtain
    \begin{equation}
        I_{12} = \ft{L_{1}O}(\vv{0}) + \ft{L_{1}O}(\vv{0}) + 2 \ft{L_{12}O}(\vv{0}) \Re \left( V_{12}  \right)
        \label{Eq:Interferogram4}
    ,\end{equation}
where $V_{12}$ is the visibility of the object that is apodised by the interferometric lobe.
When we assume that the photometric lobes are identical and equal to $L$ in the two beams, then the measured visibility writes
    \begin{equation}
        V = \frac{\ft{L_{12}O}(\vv{0})}{\ft{LO}(\vv{0})} V_{12} 
        \label{Eq:Measured_visibility}
    ,\end{equation}
and the visibility multiplicative bias is therefore the coherent-to-uncoherent photometry ratio.
As a consequence, the  visibilities are underestimated, leading to values lower than 1 even at very short baselines.
This effect is shown in Fig.~\ref{Fig:max_contrast_quiet}, where it has been studied for different separations between Sgr~A$^{\star}$ and S2, with the former in different states of brightness (bright flare, moderate flare, and quiescent state).
We have assumed that the fibres are perfectly centred on Sgr~A$^{\star}$ in the two beams of the interferometer.
The effect is almost negligible for a separation of 11.1\,mas up to a residual tip-tilt amplitude of 20\,mas because the contrast loss is 1.1\% in the quiescent state.
In all other cases, however, for larger separations, the contrast loss is much higher than the accuracy of the instrument.
It is close to 20\% at 20\,mas residual tip-tilt for a separation of 59.8\,mas in the quiescent state.
A correction has to be applied to visibilities if the rms tip-tilt can be estimated or the bias has to be modelled to fit the data. In either case, it is necessary to assess the effect of residual tip-tilt on lobes either through a calibration of the lobes or by measuring the tip-tilt variance. The non-stationarity of seeing could be a problem here, which would need to be addressed in a separate study.

\section{Conclusion}
\label{Sec:6}

The field of view of an interferometer is defined as the field in which sources contribute to spatial coherence or interferences.
It is constrained by various parameters that range from hardware limitations to spectral resolutions and (u,v) plane sampling.
Here, we focused on the special case of single-mode interferometers where the maximum field of view is set by the fibre mode.
We explored for the first time the influence of partial correction in adaptive optics on the maximum field of view of single-mode interferometers at optical wavelengths in the long-exposure regime.
This is a new regime opened by the GRAVITY instrument of the VLTI.
Past studies were made by \citet{Guyon2002} in the short-exposure regime with an application to binaries and by \citet{Mege2002}, but were restricted to the case without phase correction.
Like these authors, we defined photometric and interferometric lobes, as is also done for radio interferometers.
We also found that the two lobes are not identical, except in absence of turbulence or when the distance between the pupils of the interferometer is much larger than the outer scale of turbulence.
In the turbulent case without correction, the spatial distribution of the interferometric lobe is identical to the aberration-free case.
In these two cases, the maximum field of view of the interferometer matches the diffraction limit of a single aperture.
The situation is different when turbulent aberrations are partially corrected for.
We have studied the particular case where the wavefront residues are dominated by tip-tilt.
We have shown that the field of view is widened by tip-tilt residues, with a specific application to the Galactic centre, where stars at more than one diffraction limit away from the centre can contribute to the interferometric signal.
Although the widening of the field of view remains modest for tip-tilt residues of 10\,mas per axis, it dramatically changes the measured visibility when the central object is much fainter than the surrounding sources.
In practice, the field of view can be more than doubled.
We have also shown that the field of view can be extended by generating dynamical tip-tilt aberrations. 
Because the photometric and interferometric lobes do not match under turbulent conditions, we have discussed the effect of this issue on the calibration of visibilities when tip-tilt residuals dominate and have shown that the effect can be calibrated if the rms residual tip-tilt is known.

Last, we focused this paper on the amplitude of the photometric and interferometric lobes.
However, the phase of the single-mode interferometric lobes also plays an important role by potentially affecting the phase of the interferometric observables.
Intuitively, the more completely the observed object fills the interferometric lobe, the more its imaging and/or astrometry is affected.
We leave this investigation to a forthcoming paper.

\begin{table*}[!ht]
  \centering
  \bgroup
  \def\arraystretch{1.5}
  \begin{tabular}{c|c}
    \hline
    Single-mode flux & Single-mode coherence \\
    \hline
    \hline
      $I          =\int{[ P_ \varocircle    \otimes P_ \varocircle    ](\vv{u}) \; V(\vv{u}               ) \ud\vv{v}}$ &
      $\gamma_{12}=\int{[ P_{\varocircle,1} \otimes P_{\varocircle,2} ](\vv{u}) \; V(\vv{u}+\vv{B}/\lambda) \ud\vv{v}}$ \\
    \hline
      $L(\vv\alpha)     =\invft{P_ \varocircle    \otimes P_ \varocircle   }(\vv\alpha)$ &
      $L_{12}(\vv\alpha)=\invft{P_{\varocircle,1} \otimes P_{\varocircle,2}}(\vv\alpha)$ \\
    \hline
      $I          =\ft{L     O}(\vv{0             })$ &
      $\gamma_{12}=\ft{L_{12}O}(\vv{\vv{B}/\lambda})$ \\
    \hline
  \end{tabular}
  \egroup
  \caption{Summary of the flux and coherence quantities measured with single-mode telescope arrays.}
\end{table*}


\begin{acknowledgements}
The authors are grateful to the referee for the very useful interaction that helped us to improve the paper. This work was supported by Paris Observatory, by the Programme National Cosmologie et Galaxies (PNCG) of CNRS/INSU with INP and IN2P3, co-funded by CEA and CNES and by the Programme National GRAM of CNRS/INSU with INP and IN2P3, co-funded by CNES.
\end{acknowledgements}

\bibliographystyle{aa}
\bibliography{GRAVITY_fov}

\end{document}